\documentclass[iop]{emulateapj}
\newcommand{\kms}{{\,km\,s$^{-1}$}}

\slugcomment{}
\shorttitle{Far-infrared spectroscopy of ULIRGs}
\shortauthors{Farrah et al}

\begin{document}

\title{Far-infrared Fine-Structure Line Diagnostics of Ultraluminous Infrared Galaxies}

\author{D. Farrah\altaffilmark{1}}
\altaffiltext{1}{Department of Physics, Virginia Tech, VA 24061, USA} 

\author{V. Lebouteiller\altaffilmark{2,3}}
\altaffiltext{2}{Cornell University, Space Sciences Building, Ithaca, NY 14853, USA} 
\altaffiltext{3}{CEA-Saclay, F-91191 Gif-sur-Yvette,  France} 

\author{H. W. W. Spoon\altaffilmark{2}}

\author{J. Bernard-Salas\altaffilmark{4}}
\altaffiltext{4}{Dept. of Physics \& Astronomy, The Open University, MK7 6AA, UK.} 

\author{C. Pearson\altaffilmark{4,5}}
\altaffiltext{5}{Rutherford Appleton Laboratory, Harwell, Oxford, OX11 0QX, UK.} 

\author{D. Rigopoulou\altaffilmark{6,5}}
\altaffiltext{6}{Denys Wilkinson Building, University of Oxford, Keble Rd, Oxford OX1 3RH, UK}

\author{H. A. Smith\altaffilmark{7}}
\altaffiltext{7}{Harvard-Smithsonian Center for Astrophysics, Cambridge, MA 02138, USA} 

\author{E. Gonz\'alez-Alfonso\altaffilmark{8}}
\altaffiltext{8}{Universidad de Alcal\'a, Departamento de F\'{\i}sica y Matem\'aticas, Campus Universitario, Madrid, Spain}

\author{D. L. Clements\altaffilmark{9}}
\altaffiltext{9}{Physics Department, Imperial College London, London, SW7 2AZ, UK} 

\author{A. Efstathiou\altaffilmark{10}}
\altaffiltext{10}{School of Sciences, European University Cyprus, Diogenes Street, Engomi, 1516 Nicosia, Cyprus} 

\author{D. Cormier\altaffilmark{11}}
\altaffiltext{11}{Institut f{\"u}r theoretische Astrophysik, Zentrum f{\"u}r Astronomie der Universit{\"a}t Heidelberg, Heidelberg, Germany}

\author{J. Afonso\altaffilmark{12,13}}
\altaffiltext{12}{Centro de Astronomia e Astrof\'{\i}sica da Universidade de Lisboa, Observat\'{o}rio Astron\'{o}mico de Lisboa, Lisbon, Portugal} 
\altaffiltext{13}{Department of Physics, University of Lisbon, 1749-016 Lisbon, Portugal} 

\author{S. M. Petty\altaffilmark{1}}

\author{K. Harris\altaffilmark{1}}

\author{P. Hurley\altaffilmark{14}}
\altaffiltext{14}{University of Sussex, Brighton BN1 9QH, UK}

\author{C. Borys\altaffilmark{15}}
\altaffiltext{15}{Infrared Processing and Analysis Center, MS220-6, California Institute of Technology, Pasadena, CA 91125, USA}

\author{A. Verma\altaffilmark{6}}

\author{A. Cooray\altaffilmark{16}}
\altaffiltext{16}{Department of Physics \& Astronomy, University of California, Irvine,CA 92697, USA}

\author{V. Salvatelli\altaffilmark{17,16}}
\altaffiltext{17}{Physics Department and INFN, Universit\`a di Roma ``La Sapienza'', Ple Aldo Moro 2, 00185, Rome, Italy}

\begin{abstract}
We present Herschel observations of six fine-structure lines in 25 Ultraluminous Infrared Galaxies at $z<0.27$. The lines, [\ion{O}{3}]52$\mu$m, [\ion{N}{3}]57$\mu$m, [\ion{O}{1}]63$\mu$m, [\ion{N}{2}]122$\mu$m, [\ion{O}{1}]145$\mu$m, and [\ion{C}{2}]158$\mu$m, are mostly single gaussians with widths $<$600 km s$^{-1}$ and luminosities of $10^7 - 10^9$L$_{\odot}$. There are deficits in the [\ion{O}{1}]63/L$_{\rm{IR}}$, [\ion{N}{2}]/L$_{\rm{IR}}$, [\ion{O}{1}]145/L$_{\rm{IR}}$, and [\ion{C}{2}]/L$_{\rm{IR}}$ ratios compared to lower luminosity systems. The majority of the line deficits are consistent with dustier \ion{H}{2} regions, but part of the [\ion{C}{2}] deficit may arise from an additional mechanism, plausibly charged dust grains. This is consistent with some of the [\ion{C}{2}] originating from PDRs or the ISM. We derive relations between far-IR line luminosities and both IR luminosity and star formation rate. We find that [\ion{N}{2}] and both [\ion{O}{1}] lines are good tracers of IR luminosity and star formation rate. In contrast, [\ion{C}{2}] is a poor tracer of IR luminosity and star formation rate, and does not improve as a tracer of either quantity if the [\ion{C}{2}] deficit is accounted for. The continuum luminosity densities also correlate with IR luminosity and star formation rate. We derive ranges for the gas density and ultraviolet radiation intensity of $10^{1}<n<10^{2.5}$ and $10^{2.2}<G_{0}<10^{3.6}$, respectively. These ranges depend on optical type, the importance of star formation, and merger stage. We do not find relationships between far-IR line properties and several other parameters; AGN activity, merger stage, mid-IR excitation, and SMBH mass. We conclude that these far-IR lines arise from gas heated by starlight, and that they are not strongly influenced by AGN activity. 
\end{abstract}


\section{Introduction}
Ultraluminous Infrared Galaxies (ULIRGS, objects with L$_{\rm{IR}}>10^{12}$L$_{\odot}$, \citealt{san96,lfs06}) are a cosmologically important population whose nature changes substantially with redshift. At $z<0.3$ ULIRGs are rare \citep[e.g.][]{sne91,vacc10}, with less than one per $\sim$hundred square degrees. They are invariably mergers between approximately equal mass galaxies \citep{clem96,sur00,cui01,far01,bus02,vei02,vei06}. Evidence suggests that their IR emission arises mainly from high rates of star formation \citep{gen98,tra01,fra03,nard10,wang11}, though of order half also contain a luminous AGN \citep{rig99,far03,ima07,veg08,nris11}. The AGN in ULIRGs may become more important with increasing IR luminosity, and advancing merger stage \citep{teng10,yuan10,stier13}, and sometimes initiate powerful outflows \citep{spo08,fisch10,rup11,sturm11,west12,rod13}. A small fraction of (low-redshift) ULIRGs become optical QSOs \citep{tac02,kaw06,kaw07,far07b,meng10,hou11} and a large fraction end up as early-type galaxies \citep{gen01,das06,roth13,wang13}. 

\begin{deluxetable*}{lccccccc}
\tabletypesize{\scriptsize}
\tablecolumns{9}
\tablewidth{0pc}
\tablecaption{The Sample.}
\tablehead{
\colhead{Galaxy}&\colhead{RA (J2000)}&\colhead{Dec}&\colhead{Redshift}&\colhead{L$_{\rm{IR}}$\tablenotemark{a}}&\colhead{Opt. Class}&\colhead{Stage\tablenotemark{b}}&\colhead{SMBH Mass\tablenotemark{c}}
}
\startdata
\objectname{IRAS 00188-0856} & 00 21 26.5 & -08 39 26.3 & 0.128 & 12.39 & LINER & V    & --  \\   
\objectname{IRAS 00397-1312} & 00 42 15.5 & -12 56 02.8 & 0.262 & 12.90 & HII   & V    & 0.11 \\  
\objectname{IRAS 01003-2238} & 01 02 50.0 & -22 21 57.5 & 0.118 & 12.32 & HII   & V    & 0.25 \\  
\objectname{Mrk 1014}        & 01 59 50.2 & +00 23 40.6 & 0.163 & 12.62 & Sy1   & IIIb & 1.35 \\  
\objectname{IRAS 03158+4227} & 03 19 12.4 & +42 38 28.0 & 0.134 & 12.63 & Sy2   & IIIa & -- \\    
\objectname{IRAS 03521+0028} & 03 54 42.1 & +00 37 03.4 & 0.152 & 12.52 & LINER & IIIb & -- \\    
\objectname{IRAS 06035-7102} & 06 02 54.0 & -71 03 10.2 & 0.079 & 12.22 & HII   & IIIa & 0.09 \\  
\objectname{IRAS 06206-6315} & 06 21 01.2 & -63 17 23.5 & 0.092 & 12.23 & Sy2   & IIIb & -- \\    
\objectname{IRAS 07598+6508} & 08 04 33.1 & +64 59 48.6 & 0.148 & 12.50 & Sy1   & IVb  & 1.48 \\  
\objectname{IRAS 08311-2459} & 08 33 20.6 & -25 09 33.7 & 0.100 & 12.50 & Sy1   & IVa  & -- \\    
\objectname{IRAS 10378+1109} & 10 40 29.2 & +10 53 18.3 & 0.136 & 12.31 & LINER & IVb  & 0.10 \\  
\objectname{IRAS 11095-0238} & 11 12 03.4 & +02 04 22.4 & 0.107 & 12.28 & LINER & IIIb & 0.35 \\  
\objectname{IRAS 12071-0444} & 12 09 45.1 & -05 01 13.9 & 0.128 & 12.41 & Sy2   & IIIb & 0.30 \\  
\objectname{3C 273}          & 12 29 06.7 & +02 03 08.6 & 0.158 & 12.83 & Sy1   & V    & 24.2 \\  
\objectname{Mrk 231}         & 12 56 14.2 & +56 52 25.2 & 0.042 & 12.55 & Sy1   & IVb  & 0.17 \\  
\objectname{IRAS 13451+1232} & 13 47 33.3 & +12 17 24.2 & 0.121 & 12.32 & Sy2   & IIIb & 0.53 \\  
\objectname{Mrk 463}         & 13 56 02.9 & +18 22 19.1 & 0.051 & 11.79 & Sy2   & IIIb & 0.65 \\  
\objectname{IRAS 15462-0450} & 15 48 56.8 & -04 59 33.6 & 0.100 & 12.24 & Sy1   & IVb  & 0.69 \\  
\objectname{IRAS 16090-0139} & 16 11 40.5 & -01 47 05.6 & 0.134 & 12.55 & LINER & IVa  & --   \\  
\objectname{IRAS 19254-7245} & 19 31 21.6 & -72 39 22.0 & 0.063 & 12.09 & Sy2   & IIIb & 0.79 \\  
\objectname{IRAS 20087-0308} & 20 11 23.9 & -02 59 50.7 & 0.106 & 12.42 & LINER & IVa  & 1.94 \\  
\objectname{IRAS 20100-4156} & 20 13 29.5 & -41 47 34.9 & 0.130 & 12.67 & HII   & IIIb & --   \\  
\objectname{IRAS 20414-1651} & 20 44 18.2 & -16 40 16.2 & 0.087 & 12.22 & HII   & IVb  & 1.03 \\  
\objectname{IRAS 23230-6926} & 23 26 03.6 & -69 10 18.8 & 0.107 & 12.37 & LINER & IVa  & 0.35 \\  
\objectname{IRAS 23253-5415} & 23 28 06.1 & -53 58 31.0 & 0.130 & 12.36 & HII   & IVa  & --   \\
\enddata  
\tablecomments{Positions, redshifts and optical spectral classifications are taken from \citealt{sau00}.}
\tablenotetext{a}{ Derived by combining the Spitzer IRS+MIPS data with all available far-IR photometry for each object, integrating under the IRS spectrum while spline-fitting to the longer wavelength data. See Borys et al, in preparation, for details. The errors on the luminosities are approximately 20\% in all cases.} 
\tablenotetext{b}{ Merger stage classification \citep{vei02}, updated using higher resolution imaging where appropriate \citep{rig99,meus01,far01,bus02,vei06}. IIIa: Premerger with separation $>10$Kpc, IIIb: Premerger with separation $<10$Kpc, IVa: Diffuse merger (prominent tidal features, but only one nucleus), IVb: Compact merger, V: Undisturbed.}
\tablenotetext{c}{ SMBH mass, in units of $10^{8}$M$_{\odot}$ \citep{zhe02,das06,gree07,zha08,vei09}.}\label{sample}
\end{deluxetable*}

Over $0.3\lesssim z <1$ the number of ULIRGs rises rapidly \citep[e.g.][]{lefl05}, reaching a density on the sky of several hundred per square degree at $z\gtrsim1$ \citep{rr97,dol,bor,mor,aus10,goto11}. The fraction of $z\gtrsim1$ ULIRGs that are starburst dominated mergers is high \citep{far02b,cha03,sma04,tak06,bor06,val07,ber07,bri07,lon09,hua09,magne12,lofaro13,john13} but the merger fraction may be lower than locally (\citealt{mel08,kart10,drap12}, but see also \citealt{xu12}). High redshift ULIRGs may also have a wider range in dust temperature \citep{mag10,rr10,sym11,sym13,bri13} and SED shapes \citep{far08,saj12,nordon12}, and a greater star formation efficiency \citep{igl07,comb11,comb13,hanami12,geach13} compared to local examples. 

Determining why the number and properties of ULIRGs change so markedly with redshift may provide insight into the history of stellar and SMBH mass assembly in $\gtrsim L^{*}$ galaxies. ULIRGs at $z<0.3$ are central to this endeavour, as they establish a baseline from which to measure evolution with redshift in the ULIRG population. The far-infrared ($\simeq50-500\mu$m) is a powerful tool for studying ULIRGs, as demonstrated by the Infrared Space Observatory \citep[ISO, e.g.][]{fisch99,negi01,luh03,spin05,brau08}. The {\itshape Herschel} Space Observatory \citep{pilb10} offers dramatic advances in far-infrared observing capability over ISO. Its instruments, the Photodetector Array Camera and Spectrometer (PACS, \citealt{pogl10}), Spectral and Photometric Imaging REceiver (SPIRE, \citealt{griff10}) and Heterodyne Instrument for the Far Infrared \citep{deg10} can observe wavelength ranges that are inaccessible from the ground, and have improved sensitivity and resolution over previous space-based facilities. 

\begin{deluxetable*}{lccccccc}
\tabletypesize{\scriptsize}
\tablecolumns{8}
\tablewidth{0pc}
\tablecaption{Properties of the lines observed}
\tablehead{
\colhead{Line}&\colhead{Wavelength}&\colhead{IP$_1$}&\colhead{IP$_2$}&\colhead{Configuration}    &\colhead{$n_{cr,e}$}&\colhead{$n_{cr,H}$}&\colhead{T$_{exc}$}  \\
\colhead{}&\colhead{$\mu$m}&\colhead{eV}&\colhead{eV}&\colhead{}&\colhead{cm$^{-3}$}&\colhead{cm$^{-3}$}&\colhead{K}   
}
\startdata
\mbox{[O III]} & 51.815            & 35.12          & 54.93          & $^{3}P_{2}-^{3}P_{1}$     & $\simeq3500$          &                       & $441$ \\ 
\mbox{[N III]} & 57.317            & 29.60          & 47.45          & $^{2}P_{3/2}-^{2}P_{1/2}$ & $\simeq3000$          &                       & $251$ \\ 
\mbox{[O I]}   & 63.184            & --             & 13.62          & $^{3}P_{1}-^{3}P_{2}$     & $\simeq2.8\times10^5$ & $\simeq2.5\times10^5$ & 228   \\
\mbox{[N II]}  & 121.898           & 14.53          & 29.60          & $^{3}P_{2}-^{3}P_{1}$     & $\simeq400$           &                       & 188   \\
\mbox{[O I]}   & 145.525           & --             & 13.62          & $^{3}P_{0}-^{3}P_{1}$     & $\simeq4\times10^4$   & $\simeq5\times10^4$   & 327   \\
\mbox{[C II]}  & 157.741           & 11.26          & 24.38          & $^{2}P_{3/2}-^{2}P_{1/2}$ & $\simeq40$            & $\simeq2700$          & 91    \\
\enddata  
\tablecomments{Electron and Hydrogen critical densities are given for n=500cm$^{-3}$}\label{linesobs}
\end{deluxetable*}

We have used Herschel to conduct the Herschel ULIRG Survey (HERUS), which assembles PACS and SPIRE observations of nearly all ULIRGs with a 60$\mu$m flux greater than $\sim1.7$Jy. In this paper we present observations of fine-structure lines for 24 of the sample. Analysis of the SPIRE FTS spectra is presented in Pearson et al 2013, in preparation. Observations of the OH 119\,$\mu$m and 79\,$\mu$m profiles are presented in \citealt{spo13}, while modelling of these profiles is presented in Smith et al, in preparation. Finally, a detailed study of the  ULIRG IRAS 08572+3915 is presented in \citealt{efs13}. We define infrared luminosity, L$_{\rm{IR}}$, to be the luminosity integrated over 8-1000$\mu$m in the rest frame. We quote luminosities and masses in units of Solar (L$_{\odot} = 3.839\times10^{26}$~Watts, M$_{\odot} = 1.99\times10^{30}$~Kg, respectively). We assume a spatially flat cosmology with $H_{0} = 67.3$km s$^{-1}$~Mpc $^{-1}$, $\Omega = 1$ and $\Omega_{m} = 0.315$ \citep{planck13}.

\section{Methods}

\subsection{Sample Selection}
HERUS is a photometric and spectroscopic atlas of the $z<0.27$ ULIRG population. The sample comprises all 40 ULIRGs from the IRAS PSC-z survey \citep{sau00} with 60$\mu$m fluxes greater than 2Jy, together with three randomly selected ULIRGs with lower 60$\mu$m fluxes; IRAS 00397-1312 (1.8 Jy), IRAS 07598+6508 (1.7 Jy) and IRAS 13451+1232 (1.9 Jy). All objects have been observed with the Infrared Spectrograph (IRS, \citealt{houck04}) onboard Spitzer \citep{armus07,far07,desai07}. The SHINING survey \citep{fisch10,sturm11,hail12,gonz13} obtained PACS spectroscopy for 19/43 sources, so we observed, and present here, the remaining 24 objects. We also include Mrk 231 \citep{fisch10} to give a final sample of 25 objects (Table \ref{sample}). This sample is not flux limited, but does include nearly all ULIRGs at $z<0.27$ with 60$\mu$m fluxes between 1.7~Jy and 6~Jy,  together with Mrk 231. The sample therefore gives an almost unbiased view of $z<0.3$ ULIRGs.

\subsection{Observations}
The PACS observations were performed between March 18, 2011 and April 8, 2012 (Operational Day 673-1060). The PACS integral-field spectrometer samples the spatial direction with 25 pixels and the spectral direction with 16 pixels. Each spectral pixel scans a distinct wavelength range by varying the grating angle. The combination of the 16 ranges makes the final spectrum. The resulting projection of the PACS array on the sky is a footprint of $5\times5$ spatial pixels (“spaxels”), corresponding to a $47\arcsec\times 47\arcsec$ field-of-view. The point spread function full width at half maximum (FWHM) is $\approx 9.5\arcsec$ between 55$\mu$m\ and 110$\mu$m, and increases to about $14\arcsec$ by 200$\mu$m. A spaxel at the mean redshift of the sample is $\sim3$~kpc in extent.

A single footprint observation was performed for each object as they are all smaller than the footprint size. The coordinates were chosen to place the optical centroids in the central spaxel.  We observed the sample in the following lines: [\ion{O}{3}]52$\mu$m,  [\ion{N}{3}]57$\mu$m, [\ion{O}{1}]63$\mu$m, [\ion{N}{2}]122$\mu$m, [\ion{O}{1}]145$\mu$m, and [\ion{C}{2}]158$\mu$m (Table \ref{linesobs}). Observations were done in range spectroscopy mode. We used optical narrow-line redshifts to set the central wavelengths of each range scan. For one object, IRAS\,07598+6508, the input coordinates were incorrect, placing the source near the edge of the PACS array, thus making the flux determination uncertain. We therefore substituted observations of this source from other programs. For [\ion{C}{2}], we used the dataset 1342243534 (PI Weedman), and for [\ion{N}{2}] we used the dataset 1342231959 (PI Veilleux). 

We set the wavelength range of each range scan to accomodate uncertainties such as offsets between optical and far-IR line redshifts, and asymmetric or broadened lines. The chop/nod observation mode was used, in which the source is observed by alternating between the on-source position and a clean off-source position. Since the extent of the targets is always $<1\arcmin$, the smallest throw ($\pm1.5\arcmin$) was used to reduce the effect of field-rotation between the two chop positions. Two nod positions were used in order to eliminate the telescope background emission.

The data reduction was performed in the Herschel Interactive Processing Environment (HIPE) version \texttt{8.0} \citep{ott10} using the default chop/nod pipeline script. The level one products (calibrated in flux and in wavelength, with bad pixel masks from HIPE) were then exported, and processed by the PACSman tool \citep{lebo12}.

\begin{figure*}
\begin{center}
\includegraphics[width=110mm,angle=00]{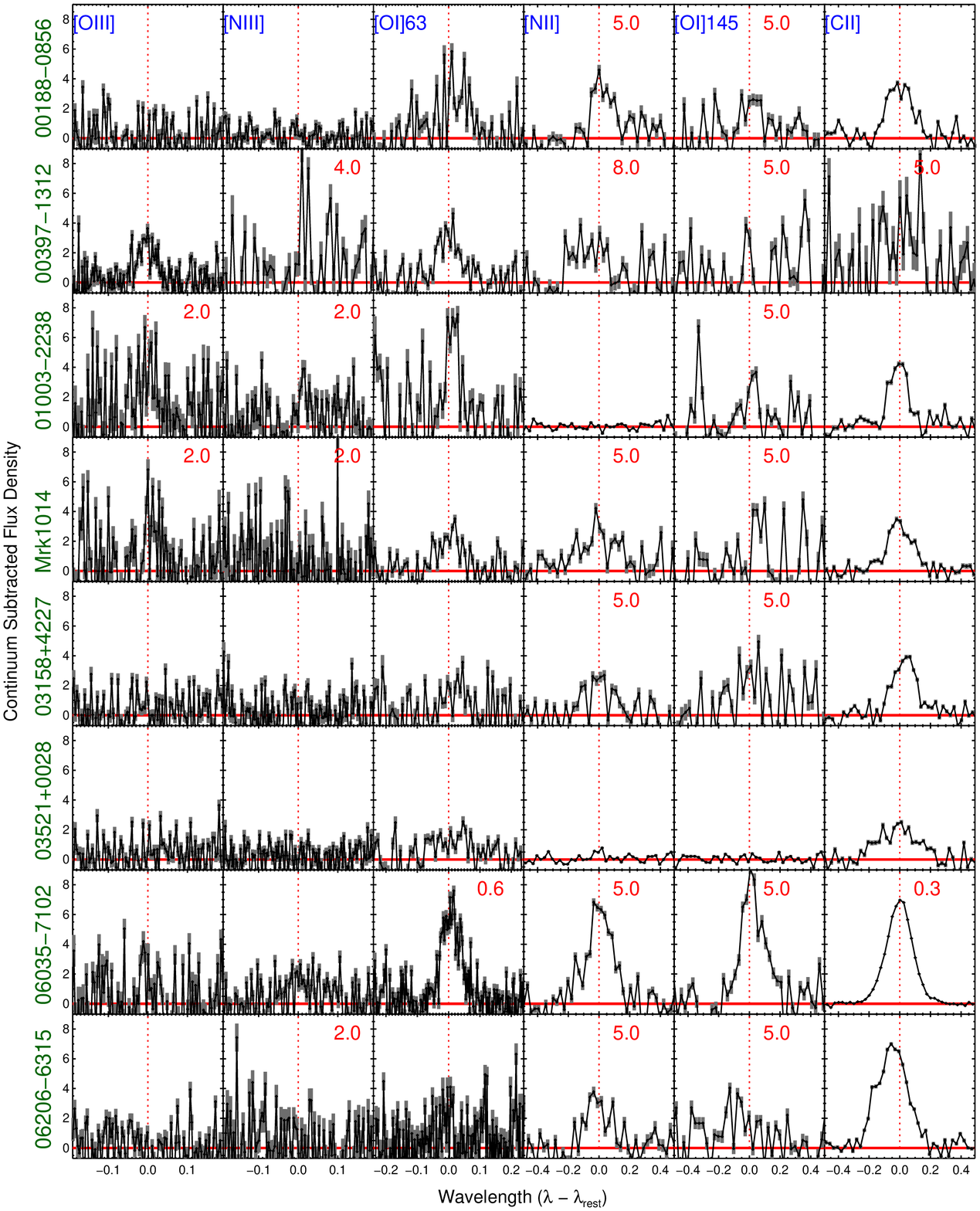}
\caption{The one-dimensional continuum-subtracted spectra for the first eight objects in Table \ref{sample}. Each row shows the spectra for one object, while each column shows the spectra for one line. The vertical red line is the (optical) redshift. If present, a red number shows the factor by which that spectrum has been scaled to fit within the $y$ axis range. We plot the rebinned spectra rather than the full data cloud, and show data only from the central spaxel.}\label{spectra1}
\end{center}
\end{figure*}

\begin{figure*}
\begin{center}
\includegraphics[width=110mm,angle=00]{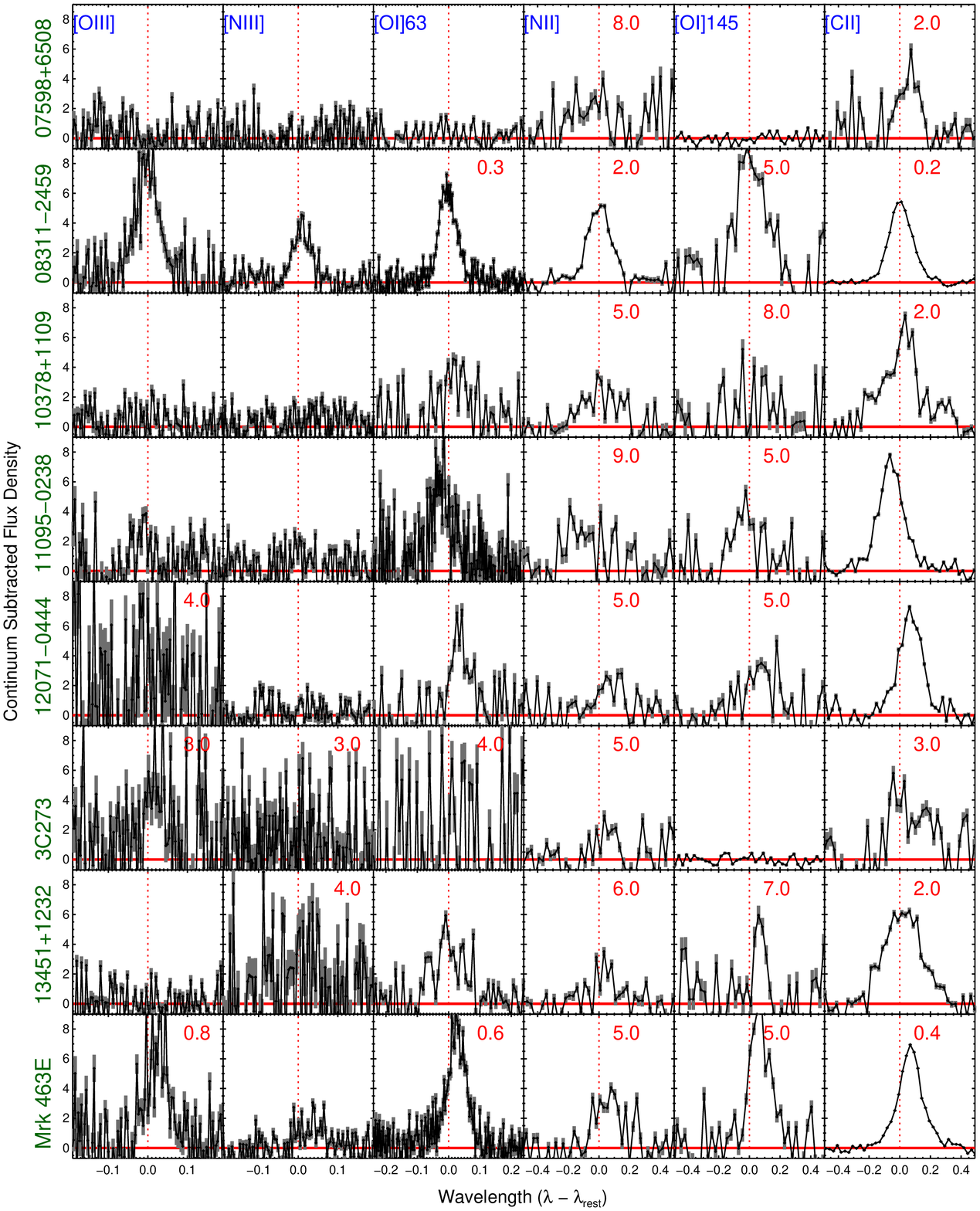}
\caption{The one-dimensional spectra of each line, for the second eight objects in Table 1, excluding Mrk 231 (see \citealt{fisch10}). The labelling of the plot is as given for Figure \ref{spectra1}.}\label{spectra2}
\end{center}
\end{figure*}

\begin{figure*}
\begin{center}
\includegraphics[width=110mm,angle=00]{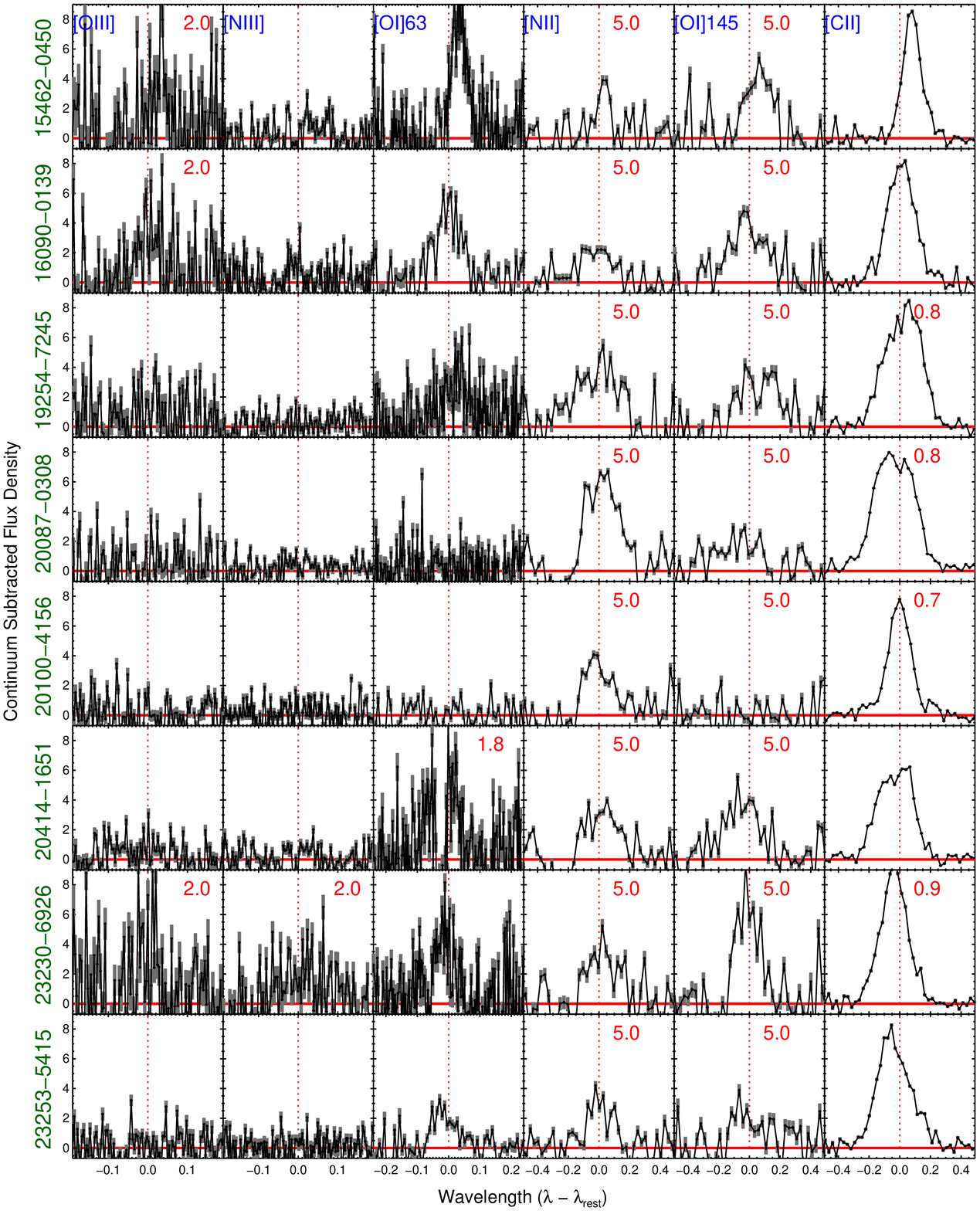}
\caption{The one-dimensional spectra of each line, for the last eight objects in Table \ref{sample}. The labelling of the plot is as given for Figure \ref{spectra1}.}\label{spectra3}
\end{center}
\end{figure*}

\subsection{Line Measurements}\label{measmeth}
The best method to determine a line flux depends on the position and morphology of the line-emitting regions within the PACS footprint. If their combined spatial extent is significantly smaller than a single spaxel, are well centered on the central spaxel, and {\itshape Herschel} maintains accurate pointing (the pointing accuracy of {\itshape Herschel} is $\sim2.5\arcsec$) then the best flux measurement is that of the central spaxel, scaled by an appropriate point-source correction. We call this method `$\mathcal{M}1$'. If these conditions are not satisfied then $\mathcal{M}1$ will give a lower limit on the flux.  

We do not know, {\itshape a priori}, the morphologies of the far-IR line emitting regions, since high spatial resolution images of this emission do not exist. Moreover, we cannot assume that the morphologies of these regions are traced reliably by emission at other wavelengths. We therefore are unable to straightforwardly distinguish between scenarios such as the source being centered and spatially extended, against the source being off-center and pointlike. Finally, we cannot assume that the morphologies of different lines in the same object are similar, since the line strengths are governed by different excitation temperatures and critical densities. 

Given these caveats, there are five further methods to determine a line flux:

\begin{itemize}

\item $\mathcal{M}2$ - fit line profiles to the central 3$\times$3 spaxels individually, sum the resulting fluxes, and apply a point-source correction (of order 15\% or less of the total flux) that accounts for the additional areal sampling of the PSF. This method is suitable if the source is spatially extended or shifted by at most a significant fraction of a spaxel. The point-source correction is wavelength-dependent, but since our range scans span wavelength ranges of order 1$\mu$m it is equivalent to either calculate the flux and then apply the correction, or apply the correction to the spectrum and then calculate the flux. 

\item $\mathcal{M}3$ - Co-add the spectra of the central 3$\times$3 spaxels, apply a point-source correction, and fit a line profile to the combined profile. This method is identical to $\mathcal{M}2$, except for additional uncertainties from combining spectra with different shaped continua. We mention this method for completeness but do not use it. 

\item $\mathcal{M}4$ - fit line profiles to those spaxels with a line detection, then sum the resulting fluxes. This is appropriate if the source is extended in any fashion, but suffers from uncertainties due to imperfect knowledge of the source morphology. 

\item $\mathcal{M}5$ - fit line profiles to every spaxel in the PACS array and sum them. The point-source correction for this method is negligible. This method will capture all of the line flux, but will have overestimated uncertainties unless the source is both bright {\itshape and} extended across at least most of the PACS field-of-view.

\item $\mathcal{M}0$ - fit a PSF, as a function of position and intensity, across the whole PACS array and adopt the best fit. This is a photometric equivalent of `optimal' extraction as described in \citealt{lebo10}.

\end{itemize}

To choose the method for the lines in each source we proceed as follows. First, we determine all six measurements for each line. If the emission is consistent with a well-centered point source, then the fluxes from all six methods will agree with each other, with larger errors for the methods that include more spaxels. We found this to be so in the majority of the sample. For these, we adopted $\mathcal{M}1$. For the others, the measurements from $\mathcal{M}2$ through $\mathcal{M}6$ were higher than $\mathcal{M}1$, but were consistent with each other. This indicates that the line emission is mostly confined to the central 3$\times$3 spaxels. We therefore discarded $\mathcal{M}5$. To avoid uncertainties in using a simulated PSF, we then used in most cases the measurements from $\mathcal{M}2$ rather than $\mathcal{M}4$ or $\mathcal{M}6$, even though the $\mathcal{M}2$ errors are larger. In most cases, for each object, the same measurement method was used for all lines. In a few cases though [\ion{C}{2}] is extended while the other lines are consistent with point sources. 

For one object, IRAS 00397-1312, [\ion{C}{2}] is redshifted such that it lies in a part of the PACS wavelength range that suffers from significant flux leakage. In this wavelength range, which spans approximately 190$\mu$m to 210$\mu$m, there is superimposed emission from the second order, at 95-110$\mu$m. Since [\ion{C}{2}] in IRAS 00397-1312 was observed with SPIRE, we substitute the SPIRE-FTS measurement for this line. 

Except for [\ion{C}{2}] in IRAS 00397-1312, line fits were performed with PACSman for each spaxel at each raster position, using all points in the data cloud. Errors were estimated from the dispersion of the cloud in each wavelength bin. The fitting function was a Gaussian profile, adjusted simultaneously with a polynomial continuum of degrees one to three. The instrument spectral resolution ranges from $\sim55$\,\kms\ to $\sim320$\kms\ depending on the band, order and wavelength. The intrinsic line broadening FWHM was calculated by calculating the quadratic difference between the observed FWHM and the instrumental FWHM, assuming gaussian profiles.

\begin{figure}
\includegraphics[width=85mm,angle=00]{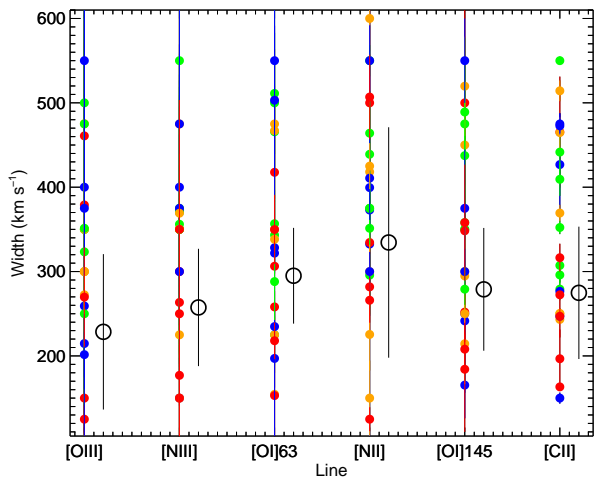}
\caption{Distribution of line widths for each line, color coded by optical class (blue: \ion{H}{2}, green: LINER, orange: Sy2, red: Sy1). See \S\ref{profileres} for discussion. Objects with errors on their widths exceeding 30$\%$ are not plotted. The open black circles show the (weighted) mean and error for each line. The line widths are corrected for instrumental broadening.}\label{linewidths}
\end{figure}

\section{Results and Analysis}
We tabulate the far-infrared line properties in Table \ref{linefluxc}, and present their profiles in Figures \ref{spectra1}, \ref{spectra2}, and \ref{spectra3}. In the following analysis we frequently compare the far-IR line properties to those of the mid-IR fine-structure lines \citep{far07}, the 6.2$\mu$m and 11.2$\mu$m Polycyclic Aromatic Hydrocarbon (PAH) features, and the 9.7$\mu$m silicate feature (Table \ref{midflux}). The PAH and silicate feature data were measured from spectra taken from the CASSIS v4 database \citep{lebo11}. The PAH luminosities were measured by integrating over 5.9-6.6$\mu$m and 10.8-11.8$\mu$m in the continuum subtracted spectra, respectively \citep{spo07}. For the silicate feature, we define its strength, $S_{Sil}$, as:

\begin{equation}\label{sildepth}
S_{Sil} = ln\left[\frac{f_{obs}}{f_{cont}}\right]
\end{equation}

\noindent where $f_{obs}$ is the observed flux at rest-frame 9.7$\mu$m and $f_{cont}$ is the continuum flux at rest-frame 9.7$\mu$m in the absence of silicate absorption, inferred from a spline fit to the continuum at 5-7$\mu$m and 14-14.5$\mu$m \citep{spo07,lev07}. A positive value corresponds to silicates in absorption.

\begin{figure*}
\begin{center}
\includegraphics[width=110mm,angle=00]{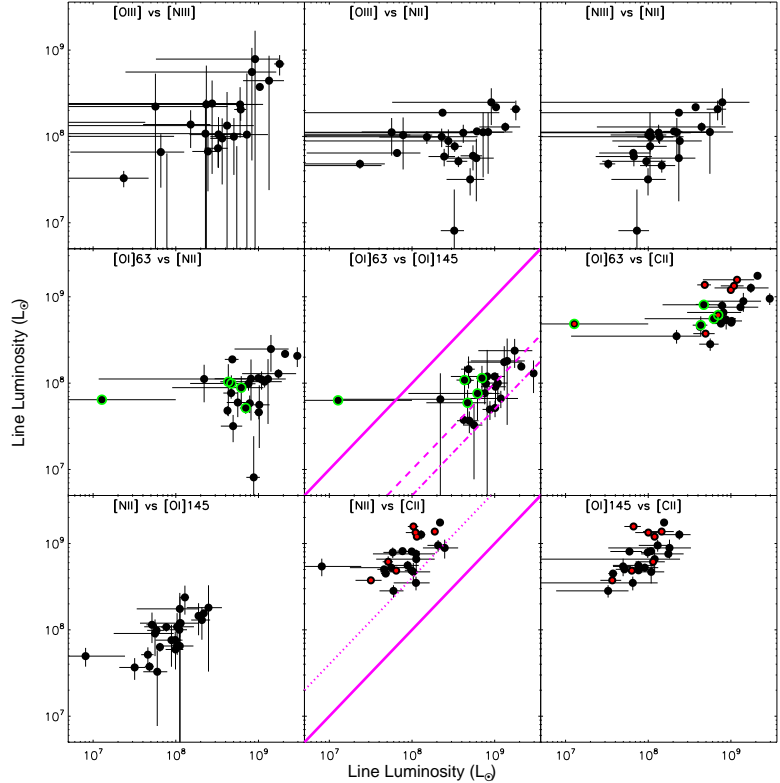}
\caption{Selected far-IR line luminosities, plotted against each other (\S\ref{lumincorrs}). The label in each panel identifies which line is plotted on the $x$ and $y$ axes, respectively. The solid, dashed and dot-dashed lines in the [\ion{O}{1}]63 vs. [\ion{O}{1}]145 plot indicate ratios of 1:1, 10:1, and 20:1, respectively. The solid and dotted lines in the [\ion{N}{2}] vs. [\ion{C}{2}] plot indicate ratios of 1:1, 1:4, respectively. Objects with green annuli may show self absorption in [\ion{O}{1}]63. Objects with red cores may show an additional component in [\ion{C}{2}].}\label{firlumvsfirlum1}
\end{center}
\end{figure*}

\subsection{Line Properties}
\subsubsection{Profile Shapes}\label{profileres}
In most cases the lines are reproducible by single gaussians with widths between 250\kms\ and 600\kms. We do not see greater widths in the higher ionization lines. We also do not see strong asymmetries, or systemic offsets in velocity (compared to the optical redshift) of any line. In a few cases the line profiles are not reproducible with single symmetric profiles. The [\ion{O}{1}]63 lines in two objects, IRAS 06206$-$6315 and IRAS 20414-1651 are consistent with significant self-absorption (see also \citealt{graci11}). There is weaker evidence for [\ion{O}{1}]63 self absorption in IRAS 00188-0856, IRAS 11095-0238, and IRAS 19254-7245. Self-absorption in [\ion{O}{1}]63 can occur when cool oxygen in foreground clouds reduces the [\ion{O}{1}]63 flux \citep{pogl96,fisch99,vast02,luh03}. The [\ion{C}{2}] profiles in IRAS 20087-0308, IRAS 20414-4651 and IRAS 23253-5415 may show subtle asymmetries. Finally, in four cases (IRAS 06035-7102, Mrk 463, IRAS 11095-0238, and IRAS 20100-4156), [\ion{C}{2}] may show an additional, broad emission component with widths between 600\kms\ and 1200\kms\. For consistency with the other lines we do not include the broad component in the line fluxes in Table \ref{linefluxc}, but discuss it instead in a future paper. 

We examine the distribution of line widths, using only the narrow components in those cases where an additional broad one exists, in Figure \ref{linewidths}. The range in widths of all six lines are consistent with each other. We see no dependence of the range in widths on optical class. For individual objects though there are sometimes substantial differences between individual line widths. In Mrk 1014 for example the [\ion{C}{2}] and [\ion{N}{2}] line widths differ by nearly a factor of two. We speculate that these differences are due to one or more of (a) differences in the critical densities of the lines, (b) that an individually unresolved broad component in one line is brighter than the equivalent component in the other line, and (c) in the case of [\ion{C}{2}] and [\ion{N}{2}], that the [\ion{N}{2}] emission arises mostly from \ion{H}{2} regions, while at least some of the [\ion{C}{2}] emission arises from photodissociation regions (PDRs) or the diffuse ISM. It is however also possible that the [\ion{N}{2}] profile is affected by the $4_{32}-4_{23}$ transition of o-H$_{2}$O at 121.72$\mu$m \citep{fisch10,gonz10}. We discuss this in \citealt{spo13}.

\begin{figure*}
\begin{center}
\includegraphics[width=120mm,angle=00]{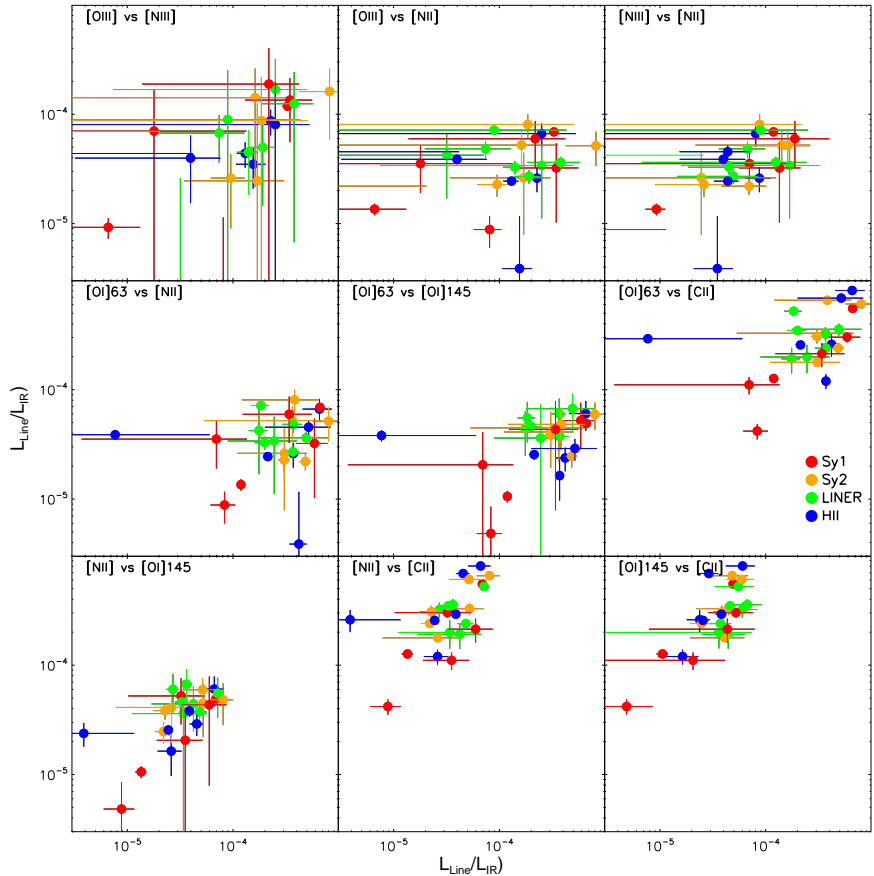}
\caption{Selected far-IR line pairs normalized by L$_{\rm{IR}}$, plotted against each other (\S\ref{lumincorrs}, see also Figure \ref{firlumvsfirlum1}).}\label{normfirlumvsfirlum1}
\end{center}
\end{figure*}

\subsubsection{Luminosities}\label{lumincorrs}
We plot luminosities of selected individual lines against each other in Figure \ref{firlumvsfirlum1}. The line luminosities range from just under $10^7$L$_{\odot}$ to  $\sim3\times10^9$L$_{\odot}$. The [\ion{C}{2}] or [\ion{O}{1}]63 lines are usually the most luminous lines, though the [\ion{O}{3}] line may be the second most luminous in 3C273 and IRAS 00397-1312. The least luminous two lines are in most cases [\ion{N}{2}] and [\ion{O}{1}]145. In some cases [\ion{N}{3}] is consistent with being the least luminous, and in a few cases (e.g. IRAS 07598+6508 IRAS 23253-5415) [\ion{O}{3}] is consistent with being the least luminous. The range in luminosities decreases approximately with increasing wavelength; the [\ion{O}{3}] and  [\ion{N}{3}] line luminosities span a factor of $\sim50$, whereas the [\ion{C}{2}] luminosity spans only a factor of five, less than the range in L$_{\rm{IR}}$ of the sample. 

To determine if the line luminosities correlate with each other, we fit the relation in Equation \ref{baserelation} (see \S\ref{linelumsdisc} for methodology). We find positive correlations in all cases, with $\beta$ values mostly between 0.5 and 1.2. In no case however is the correlation particularly strong (in terms of the S/N on $\beta$). We find that, among these six lines, the luminosity of one cannot be used to predict the luminosity of another to better than $\sim0.2$dex. If we instead plot line luminosities normalized by L$_{\rm{IR}}$ (Figure \ref{normfirlumvsfirlum1}) then the correlations do not change significantly. Neither do they depend on optical spectral type.

One object, IRAS 01003-2238, may have an unusually low [\ion{N}{2}] luminosity. The six line measurement methods for [\ion{N}{2}] in this object are in reasonable agreement, despite the low S/N. We therefore do not have a good explanation for this. The only other way in which IRAS 01003-2238 is atypical is that its optical spectrum contains Wolf-Rayet star features \citep{far05} but we do not see why this would affect the [\ion{N}{2}] luminosity. 

Finally, we comment briefly on two line ratios. First is [\ion{O}{1}]145/[\ion{O}{1}]63, which is related to the physical conditions in PDRs. This ratio (note the inverse is plotted in Figure \ref{firlumvsfirlum1}) depends on both the gas temperature and density. It decreases with decreasing gas temperature, and decreases with {\itshape increasing} gas density if the density is above the critical density. In a diffuse PDR illuminated by intense UV radiation, the expected [\ion{O}{1}] ratio lies below $\sim0.1$ (e.g., \citealt{kauf06}). From Figure 5, several of our sample have [\ion{O}{1}] ratios above 0.1. A likely reason for this is self absorption of [\ion{O}{1}]63, or a higher optical depth in [\ion{O}{1}]63 than [O I]145 (e.g., \citealt{abel07}). 

Second is the [\ion{N}{2}]/[\ion{C}{2}] ratio, which can be used to estimate the relative contribution of [\ion{C}{2}] arising in PDRs and in the ionized gas. While [\ion{N}{2}] arises almost entirely from \ion{H}{2} regions, [\ion{C}{2}] can additionally arise from PDRs and the diffuse ISM (cold neutral, warm neutral, and warm diffuse ionized). The [\ion{N}{2}]/[\ion{C}{2}] ratio can therefore quantify the contribution to [\ion{C}{2}] from the ionized gas, by comparing the observed [\ion{N}{2}]/[\ion{C}{2}] ratio with theoretical expectations from a photoionized nebula. The expected ratio in the ionized gas lies between 0.2 and 2.0 depending on density (e.g., figure 7 of \citealt{ber12}). The ratios in our sample are however nearly all below 0.25. This indicates that a significant fraction of the [\ion{C}{2}] emission originates from photodissociation regions (PDRs). A precise estimate is however hampered by lack of knowledge on the ionized gas density.

\begin{figure*}
\begin{center}
\includegraphics[width=110mm,angle=00]{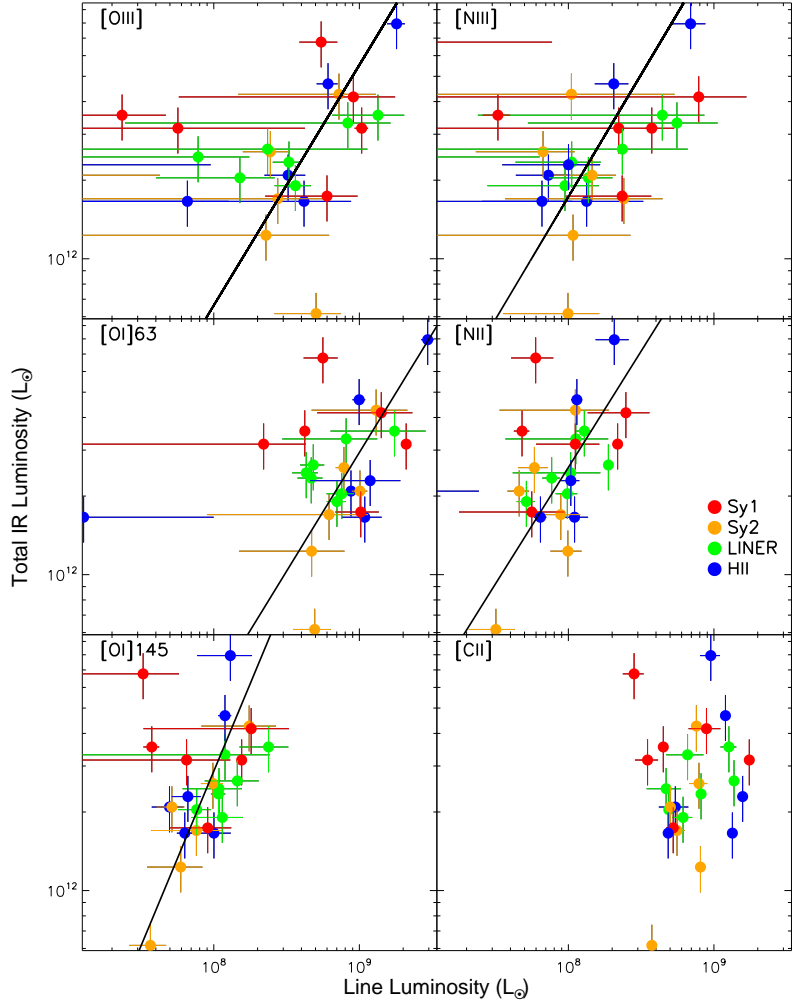}
\caption{Far-IR line luminosities vs L$_{\rm{IR}}$ (\S\ref{linelumsdisc}). The lines show the fits in Equations \ref{relations1a} to \ref{relations1f}, if the fit is significant. Here and elsewhere, we use the same $x$ and $y$ axis ranges for each line, while still showing most of the objects and all significant trends.}\label{firlinlumvslirfits}
\end{center}
\end{figure*}

\begin{figure*}
\begin{center}
\includegraphics[width=110mm,angle=00]{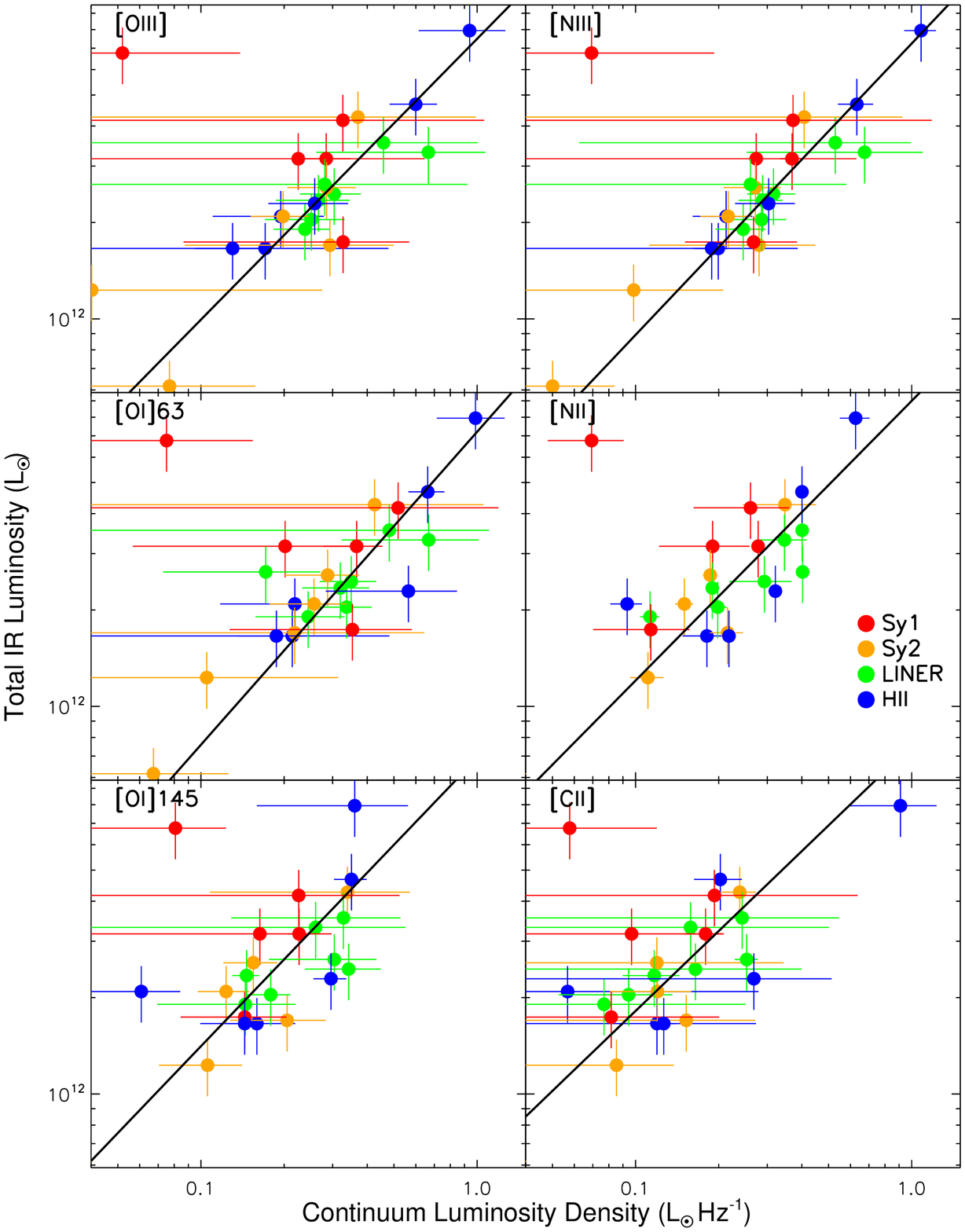}
\caption{Continuum luminosity densities near the wavelength of the indicated line, vs total IR luminosity (\S\ref{contlumslir}). The lines show fits with the Sy1s removed (Equation \ref{relations2a} through \ref{relations2f}). The Sy1 object with low flux densities near all six lines but a high IR luminosity is 3C 273.}\label{lumdensvslircolbyOpt}
\end{center}
\end{figure*}

\subsection{Infrared Luminosity}\label{irlumdisc}

\subsubsection{Line Luminosities}\label{linelumsdisc}
We compare far-IR line luminosities to L$_{\rm{IR}}$ in Figure \ref{firlinlumvslirfits}. Qualitatively, some of the line luminosities crudely correlate with L$_{\rm{IR}}$, with greater scatter among the Sy1s. To determine if correlations exist between L$_{\rm{IR}}$ and L$_{\rm{Line}}$, we assume that L$_{\rm{IR}}$ and L$_{\rm{Line}}$ are related by:

\begin{equation}\label{baserelation}
\rm{log}\left(\frac{L_{\rm{IR}}}{L_{\odot}}\right) = \alpha + \beta\rm{log}\left(\frac{L_{\rm{Line}}}{L_{\odot}}\right)
\end{equation}

\noindent where $\alpha$ and $\beta$ are free parameters. We fit this relation following the method of \citet{tremaine02}. For a correlation to exist, we require that $\beta>0$ at $>3.5\sigma$ significance. We do not claim that Equation \ref{baserelation} codifies an underlying physical relation. Neither do we claim that Equation \ref{baserelation} is the only relation that applies over the luminosity range of our sample\footnote{Indeed, in most cases through this paper where fit results are quoted, a linear ($Y= \alpha + \beta X$) or exponential ($Y = \alpha X^{\beta}$) model with appropriate choices for $\alpha$ and $\beta$ both serve equally well}. Finally, we do not consider more complex models as our data do not prefer them. 

Excluding objects with Sy1 spectra yields the following relations between L$_{\rm{IR}}$ and L$_{\rm{Line}}$:

\begin{mathletters}
\begin{eqnarray}
\rm{log}L_{\rm{IR}} & = & (4.46\pm1.77) + (0.92\pm0.20)\rm{log}(L_{\rm{[OIII]}})  \label{relations1a} \\
                    & = & (4.74\pm1.59) + (0.94\pm0.19)\rm{log}(L_{\rm{[NIII]}})  \label{relations1b} \\
                    & = & (4.28\pm1.89) + (0.91\pm0.21)\rm{log}(L_{\rm{[OI]63}})  \label{relations1c} \\
				    & = & (5.29\pm1.57) + (0.89\pm0.20)\rm{log}(L_{\rm{[NII]}})   \label{relations1d} \\
                    & = & (1.75\pm2.11) + (1.34\pm0.27)\rm{log}(L_{\rm{[OI]145}}) \label{relations1e} \\
                    & = & (6.73\pm2.44) + (0.64\pm0.27)\rm{log}(L_{\rm{[CII]}})   \label{relations1f}    
\end{eqnarray}
\end{mathletters}

\noindent which are significant (in terms of $\beta$) for all the lines except [\ion{C}{2}]. There is the caveat though that some of the relations are based on a small number of formal detections (Table \ref{linefluxc}). Including the Sy1s yields significant correlations only for [\ion{O}{3}] and [\ion{O}{1}]145, both with flatter slopes than the above.  The insignificant correlation for [\ion{C}{2}] suggests that it traces L$_{\rm{IR}}$ to an accuracy of about an order of magnitude at best. This result is consistent with previous authors, who find a crude correlation over a wider luminosity range \citep{sargs12}. 

We also investigated whether sums of the lines in Equations \ref{relations1a} to \ref{relations1f} show stronger relations with L$_{\rm{IR}}$ than individual lines. We found correlations in several cases, but none that improved significantly on those for the individual lines. 

The weaker correlations between L$_{\rm{IR}}$ and L$_{\rm{Line}}$ when Sy1s are included in the fits is consistent with far-IR lines being better tracers of L$_{\rm{IR}}$ in ULIRGs without an optical AGN. If this is true then we may see a similar result if we include in the fits only those objects with prominent PAH features, since PAHs are probably exclusively associated with star formation (see \S\ref{linessfrdisc}). We test this hypothesis by repeating the fits, but this time including only those objects with 11.2$\mu$m PAH EWs greater than 0.05$\mu$m. We chose the 11.2$\mu$m PAH because (a) it is bright, and (b) its value correlates well with the evolutionary paradigm for ULIRGs in \citet{far09}\footnote{Using the 11.2$\mu$m PAH may however lead to more bias than using the 6.2$\mu$m PAH if absorption from ices and silicate dust is primarily associated with the {\itshape background} continuum source}. We find:

\begin{mathletters}
\begin{eqnarray}
\rm{log}L_{\rm{IR}} & = & (7.01\pm1.07) + (0.63\pm0.12)\rm{log}(L_{\rm{[OIII]}})  \label{relations1g} \\
                    & = & (6.52\pm1.33) + (0.71\pm0.16)\rm{log}(L_{\rm{[NIII]}})  \label{relations1h} \\
                    & = & (6.71\pm1.30) + (0.64\pm0.15)\rm{log}(L_{\rm{[OI]63}})  \label{relations1i} \\
					& = & (6.43\pm1.51) + (0.74\pm0.19)\rm{log}(L_{\rm{[NII]}})   \label{relations1j} \\
                    & = & (2.77\pm2.12) + (1.20\pm0.26)\rm{log}(L_{\rm{[OI]145}}) \label{relations1k} \\
                    & = & (9.44\pm2.20) + (0.33\pm0.25)\rm{log}(L_{\rm{[CII]}})   \label{relations1l}   
\end{eqnarray}
\end{mathletters}

\noindent These relations are somewhat flatter than those in Equations \ref{relations1a} to \ref{relations1f}. Again the [\ion{C}{2}] line shows no significant correlation. We conclude that these far-IR lines are better tracers of L$_{\rm{IR}}$ in systems without type 1 AGN, but that it is unclear whether they are better tracers of L$_{\rm{IR}}$ in systems with more prominent star formation.

\begin{figure*}
\begin{center}
\includegraphics[width=110mm,angle=00]{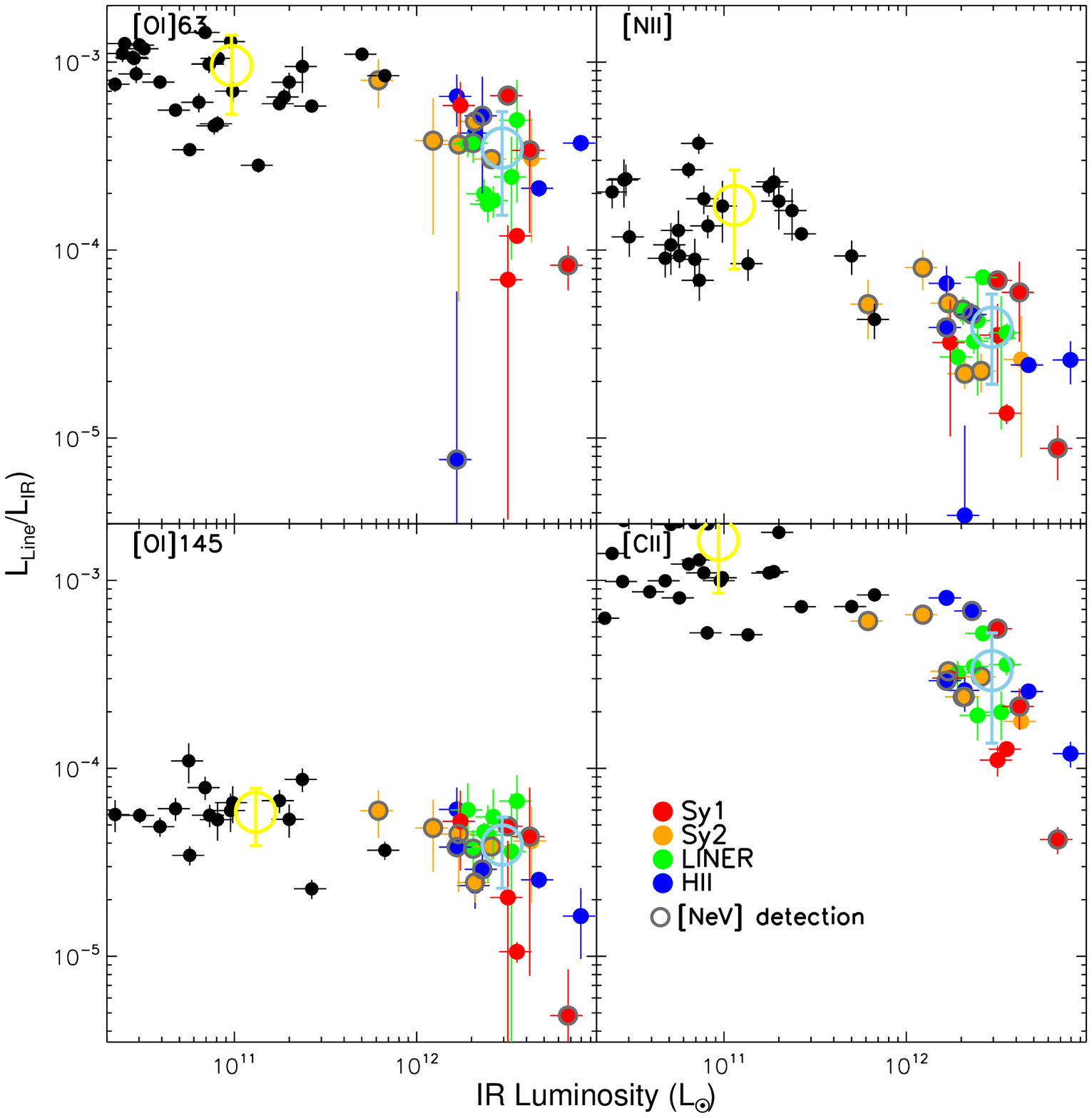}
\caption{Comparison of the [\ion{O}{1}]63/L$_{\rm{IR}}$, [\ion{N}{2}]/L$_{\rm{IR}}$, [\ion{O}{1}]145/L$_{\rm{IR}}$, and [\ion{C}{2}]/L$_{\rm{IR}}$ ratios with L$_{\rm{IR}}$ (\S\ref{defdisc}). The colored points are our sample, coded by optical class. ULIRGs with a grey annulus have a [\ion{Ne}{5}]14.32 detection. The black points are taken from \citet{brau08}. The yellow and light blue open symbols show the means and dispersions for the Brauher and our samples, respectively. The ULIRGs show a deficit compared to the lower luminosity systems in all four cases. There is however no clear dependence of the deficit on optical class, or the detection of [\ion{Ne}{5}]14.32.}\label{ciilirplots}
\end{center}
\end{figure*}

\subsubsection{Continuum Luminosities}\label{contlumslir}
We compare L$_{\rm{IR}}$ to continuum luminosity densities (in units of L$_{\odot}$ Hz$^{-1}$) near the wavelengths of the far-IR lines in Figure \ref{lumdensvslircolbyOpt}. We find a significant correlation at all six wavelengths, irrespective of whether Sy1s are included (though the continua of the Sy1s are usually not detected). For consistency with the line comparisons, we exclude the Sy1s and fit relations of the form in Equation \ref{baserelation}, obtaining:

\begin{mathletters}
\begin{eqnarray}
\rm{log}L_{\rm{IR}} & = & (12.87\pm0.09) + (0.88\pm0.17)\rm{log}(L_{\rm{52}}) \label{relations2a} \\
                    & = & (12.86\pm0.05) + (0.91\pm0.11)\rm{log}(L_{\rm{57}}) \label{relations2b} \\
                    & = & (12.86\pm0.08) + (0.98\pm0.18)\rm{log}(L_{\rm{63}}) \label{relations2c} \\
				    & = & (12.95\pm0.07) + (0.87\pm0.11)\rm{log}(L_{\rm{122}}) \label{relations2d} \\
                    & = & (13.05\pm0.11) + (0.90\pm0.16)\rm{log}(L_{\rm{145}}) \label{relations2e} \\
                    & = & (13.08\pm0.11) + (0.83\pm0.11)\rm{log}(L_{\rm{158}}) \label{relations2f}   
\end{eqnarray}
\end{mathletters}

\noindent Including the Sy1s yields comparable relations. If we instead exclude objects with PAH 11.2$\mu$m EW $\lesssim0.05$ then we obtain consistent relations.

\begin{figure*}
\begin{center}
\includegraphics[width=110mm,angle=00]{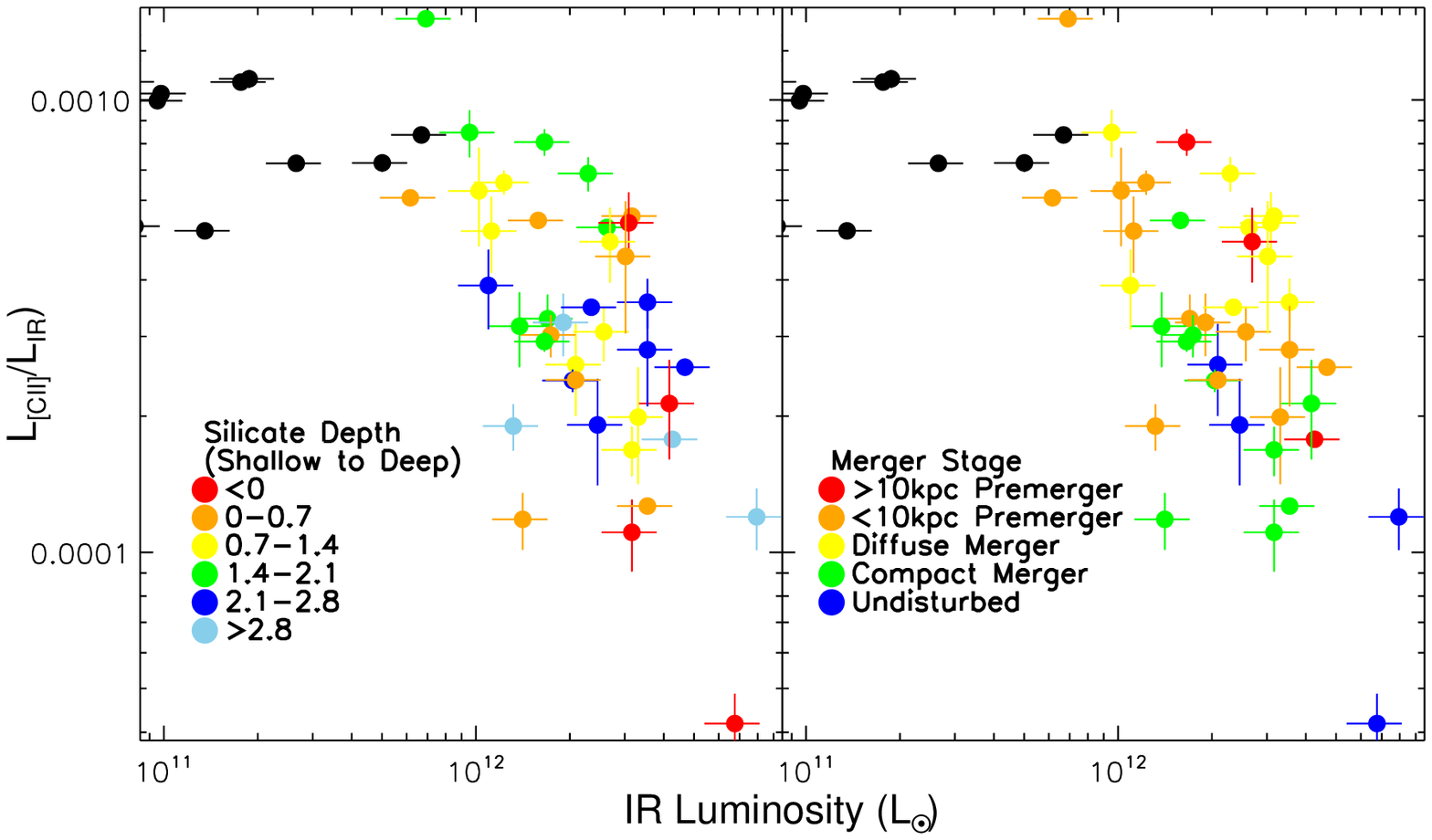} \\
\vspace{-0.7cm}
\includegraphics[width=110mm,angle=00]{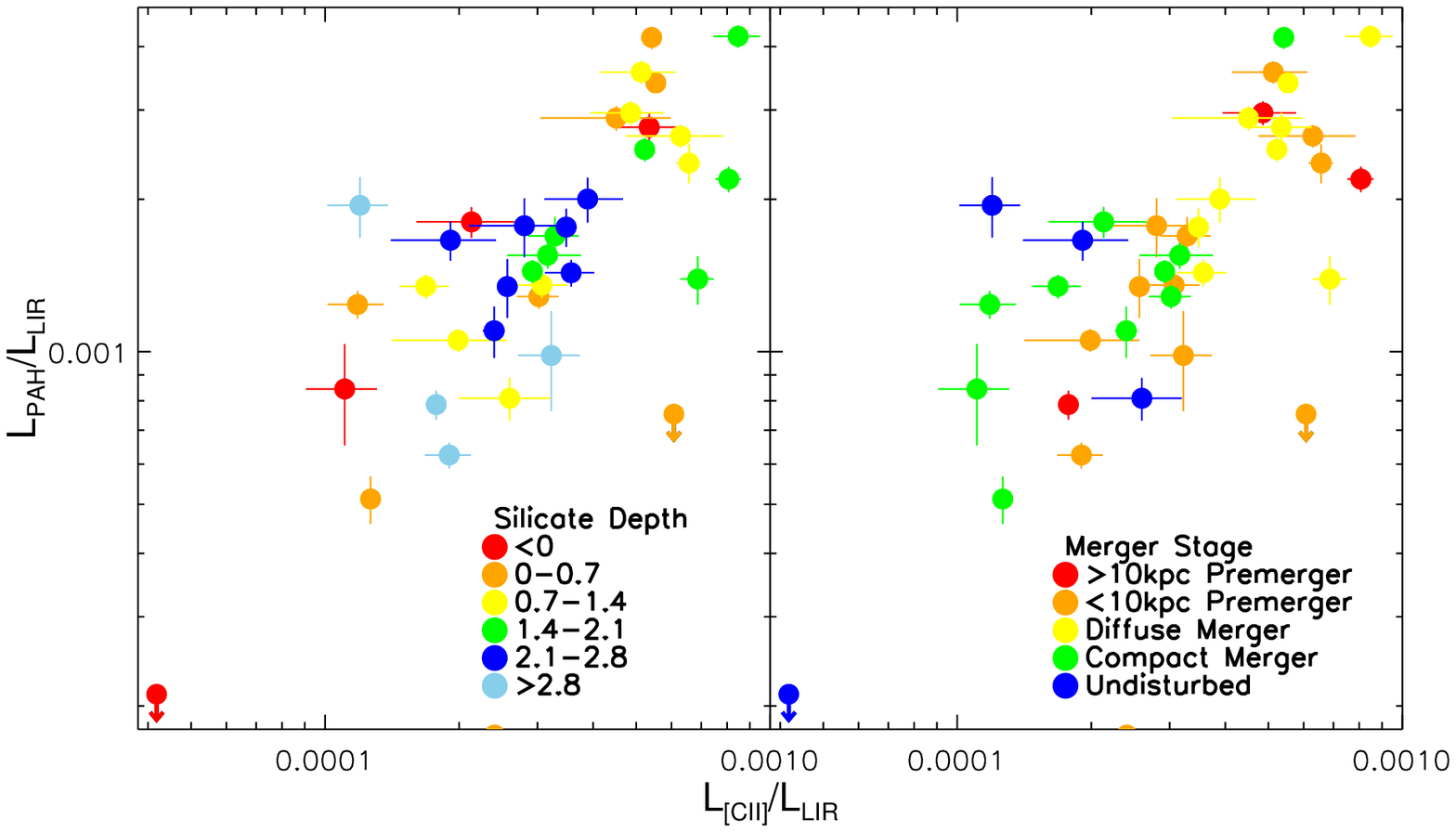}
\caption{Top row: Zoom in of the [\ion{C}{2}]/L$_{\rm{IR}}$ deficit in ULIRGs, as a function of merger stage and silicate depth (\S\ref{defdisc}). Other details are as in Figure \ref{ciilirplots}. We include in this plot additional ULIRGs with [\ion{C}{2}] detections; Arp220, Mrk 273, NGC 6240, IRAS 04103-2838, IRAS 05189-2524, IRAS 10565+2448, IRAS 12018+1941, IRAS 13342+3932, IRAS 15001+1433, IRAS 17208-0014, IRAS 20037-1547, IRAS 20100-4156, IRAS 20551-4250, IRAS 23128-5919 \citep{brau08}. Bottom row: [\ion{C}{2}]/L$_{\rm{IR}}$ plotted against L$_{\rm{PAH}}$/L$_{\rm{IR}}$.}\label{ciilirplots_si}
\end{center}
\end{figure*}

\begin{figure*}
\begin{center}
\includegraphics[width=120mm,angle=00]{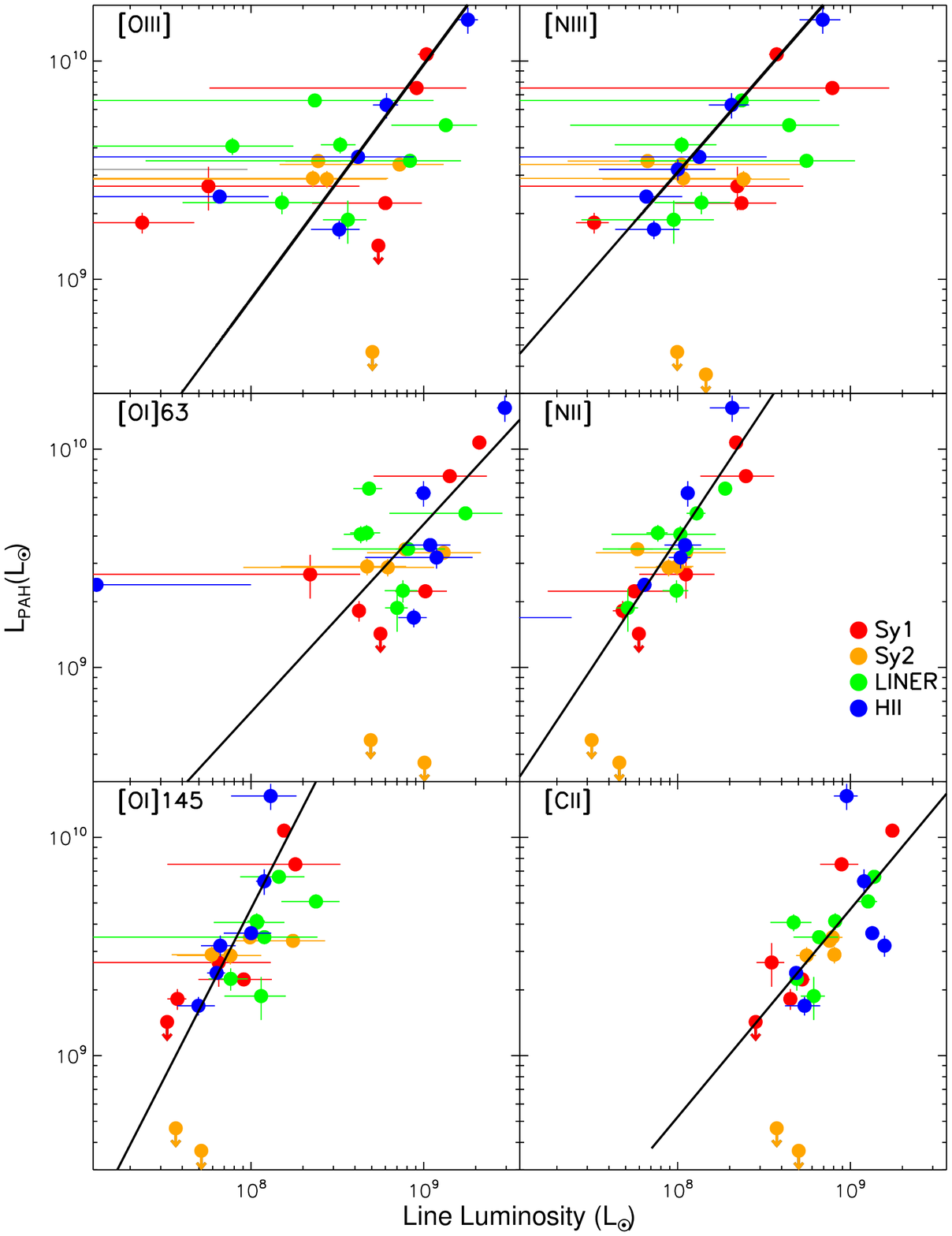}
\caption{Far-IR line luminosities vs L$_{\rm{PAH}}$ (\S\ref{linessfrdisc}). The black lines show the fits in Equation \ref{xpahsfr1a} through \ref{xpahsfr1f}.}\label{firlinevspahlumfit}
\end{center}
\end{figure*}

\subsubsection{Line Deficits}\label{defdisc}
Several far-IR lines in ULIRGs show a deficit in their L$_{\rm{Line}}$/L$_{\rm{IR}}$ ratios compared to the ratios expected from systems with lower values of L$_{\rm{IR}}$ \citep[e.g.][]{luh03}. In contrast, high redshift ULIRGs, at least for [\ion{O}{1}]63 and [\ion{C}{2}], do not show such pronounced deficits \citep{stacey10,coppin12,rig13}. We examine if the [\ion{O}{1}], [\ion{N}{2}] and [\ion{C}{2}] lines in our sample show such deficits in Figure \ref{ciilirplots} (we do not test the other lines as we lack archival data to compare to). We find a deficit in all four lines. Comparing the mean ratios at $10^{11}$L$_{\odot}$ and $10^{12.2}$L$_{\odot}$ we find differences of factors of 2.75, 4.46, 1.50, and 4.95 for [\ion{O}{1}]63, [\ion{N}{2}], [\ion{O}{1}]145, and [\ion{C}{2}], respectively. 

The four line deficits show no clear dependence on optical spectral type, the presence of an obscured AGN (as diagnosed from the detection of [\ion{Ne}{5}]14.32, see \S\ref{agnactivity}), or on PAH 11.2$\mu$m EW, for any line. There are however trends with both $S_{Sil}$ and merger stage:

\begin{itemize}

\item If $S_{Sil}\gtrsim1.4$ then the [\ion{C}{2}] and [\ion{N}{2}] deficits increase with increasing $S_{Sil}$. There is however no obvious trend of the [\ion{C}{2}] and [\ion{N}{2}] deficits with $S_{Sil}$ at $S_{Sil}\lesssim1.4$ (top left panel of Figure \ref{ciilirplots_si} \& Figure \ref{firlinevsneratio}). Conversely the [\ion{O}{1}] deficits show no trends with $S_{Sil}$.
 
\item We find no evidence that the [\ion{N}{2}] and [\ion{O}{1}] deficits depend on merger stage, but the [\ion{C}{2}] deficit is stronger, on average, in advanced mergers (classes IVb and V) than in early-stage mergers (classes IIIa through IVa, top right panel of Figure \ref{ciilirplots_si}\footnote{We also constructed the deficit plots as a function of merger stage using only those sources with $S_{Sil}<1.4$. The [\ion{C}{2}] deficit still strengthens with advancing merger stage, while no trends emerge with merger stage for the other lines.}, see also \citealt{diaz13}).

\end{itemize}

Finally, we plot in the bottom row of Figure \ref{ciilirplots_si} L$_{\rm{[CII]}}$/L$_{\rm{IR}}$ against L$_{\rm{PAH}}$/L$_{\rm{IR}}$ as a function of merger stage and $S_{Sil}$. We see in both plots consistent trends; ULIRGs in advanced mergers and with $S_{Sil}\gtrsim2$ have lower L$_{\rm{[CII]}}$/L$_{\rm{IR}}$ and L$_{\rm{PAH}}$/L$_{\rm{IR}}$ ratios, compared to ULIRGs in early-stage stage mergers and with $S_{Sil}\simeq1.4-2$.

\begin{figure*}
\begin{center}
\includegraphics[width=110mm,angle=00]{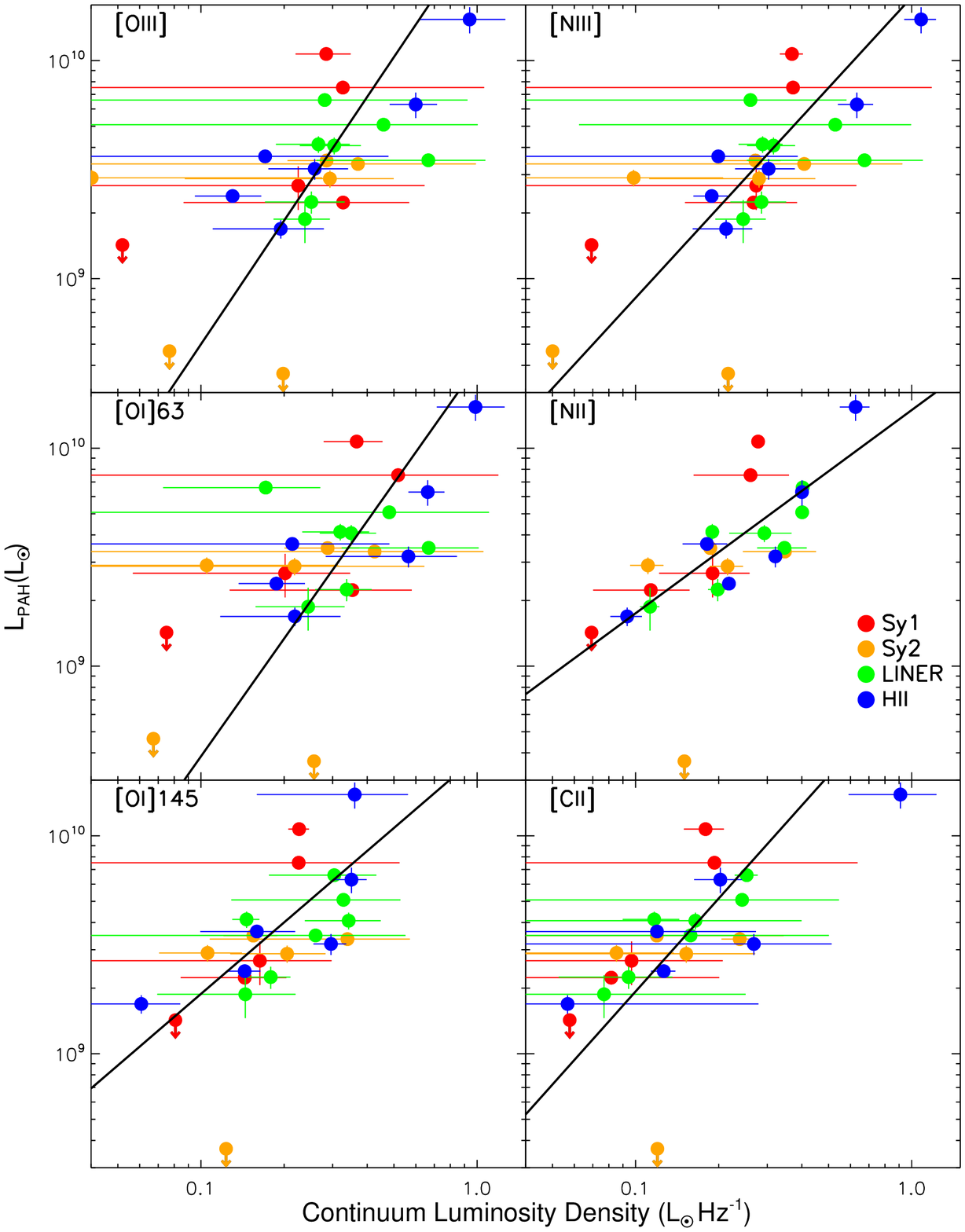}
\caption{Continuum luminosity densities at the wavelength of the indicated line, vs L$_{\rm{PAH}}$ (\S\ref{contsfrdisc}). The fits are given in Equations \ref{relations6a} to \ref{relations6f}. }\label{contlumssfr}
\end{center}
\end{figure*}

\begin{figure*}
\begin{center}
\includegraphics[width=110mm,angle=00]{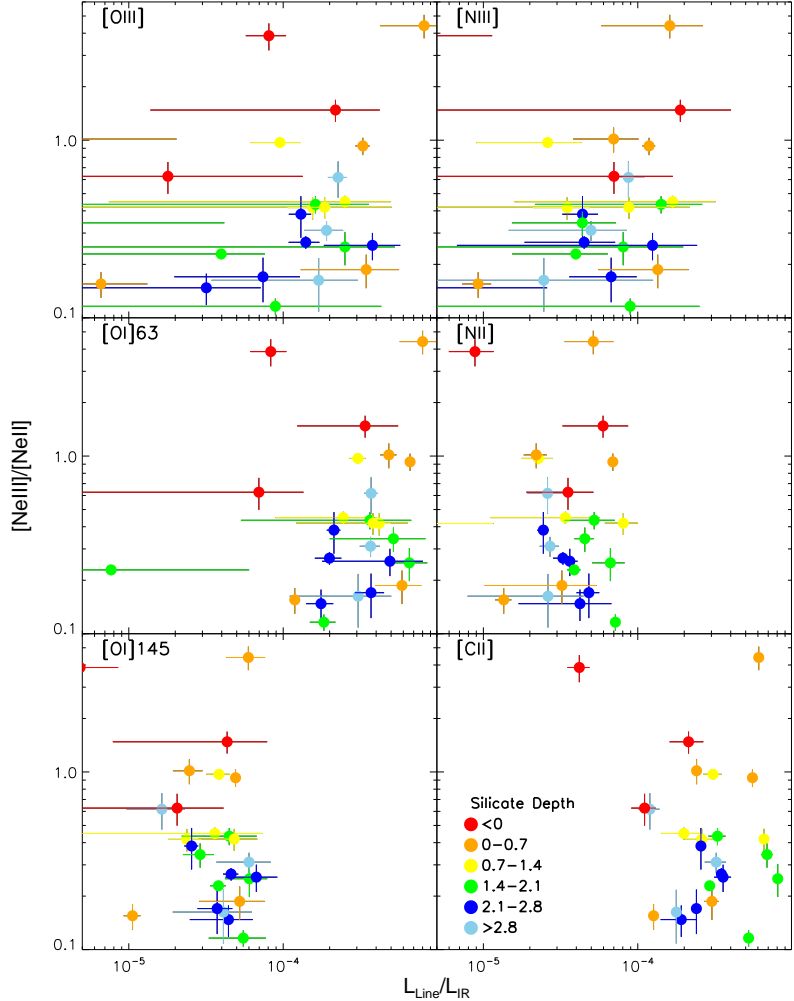}
\caption{Normalized far-IR line luminosities, plotted against [\ion{Ne}{3}]15.56/[\ion{Ne}{2}]12.81. There are no clear correlations (\S\ref{photoio}). We code the points by $S_{Sil}$ to highlight the trends described in \S\ref{defdisc}.}\label{firlinevsneratio}
\end{center}
\end{figure*}

\begin{figure*}
\begin{center}
\includegraphics[width=110mm,angle=00]{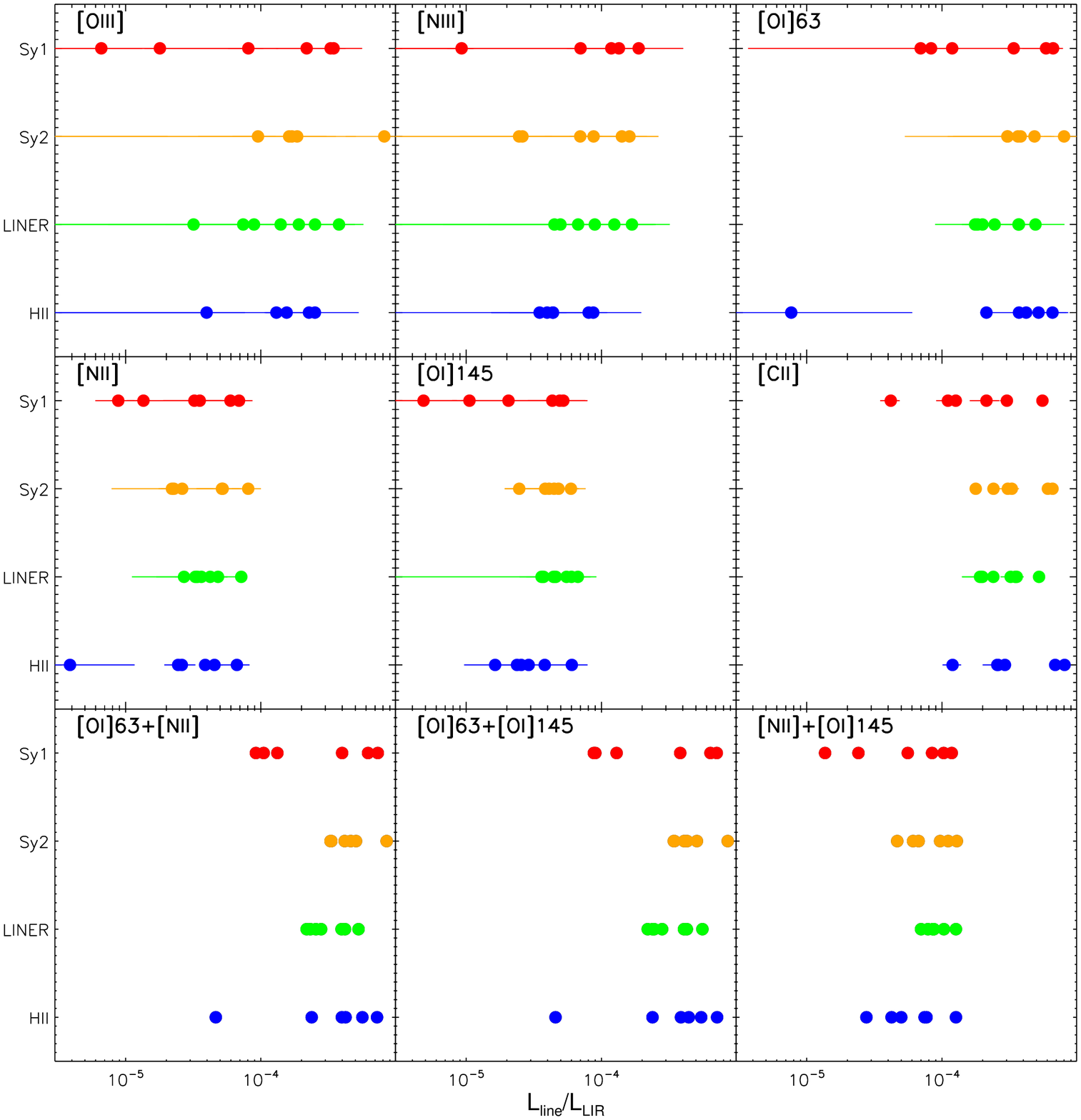}
\caption{Normalized far-IR line luminosities vs optical class (\S\ref{agnactivity}).}\label{firlinevsoptclass}
\end{center}
\end{figure*}

We do not believe that the line deficits arise from missing an asymmetric or broad component (see \S\ref{profileres}) since such components are rare. Neither do we believe the deficits arise due to self-absorption, since we see no self absorption in [\ion{C}{2}], which shows the strongest deficit. We do however see self absorption in [\ion{O}{1}]63, which shows a weaker deficit. Instead, the stronger [\ion{C}{2}] {\itshape and} [\ion{N}{2}] deficits in sources with higher $S_{Sil}$ (at $S_{Sil}\gtrsim1.4$) are consistent with the \ion{H}{2} regions in ULIRGs being dustier than \ion{H}{2} regions in lower luminosity systems. In this scenario \citep{luh03,gonz08,abel09,graci11}, a higher fraction of the UV photons are absorbed by dust rather than neutral Hydrogen, thus contributing more to L$_{\rm{IR}}$ but less to the photoionization heating of gas in the \ion{H}{2} regions, thus decreasing line emission relative to L$_{\rm{IR}}$. This mechanism would also produce a deficit in [\ion{C}{2}] and the `deficit' in the PAH emission, even if the bulk of the [\ion{C}{2}] and PAHs are in the PDRs, since there would also be fewer UV photons for photoelectric heating of the PDRs. Furthermore, this mechanism is consistent with the [\ion{O}{1}] deficits. From figure 3 of \citealt{graci11} the conditions consistent with the observed deficits of all four lines are $n_{h}\lesssim300$cm$^{-3}$ and $0.01\lesssim\langle U \rangle \lesssim 0.1$.

We propose though that dustier \ion{H}{2} regions are not the sole origin of the line deficits. This is based on three observations. First, it is puzzling that we see no strong dependence of any of the line deficits on $S_{Sil}$ when $S_{Sil}\lesssim1.4$; if the deficit arises entirely in \ion{H}{2} regions, {\itshape and} if $S_{Sil}$ is a proxy for the dust column in these \ion{H}{2} regions at $S_{Sil}\gtrsim0$, then we should see a dependence. We note though that we have only a small sample, so we could be missing a weak dependence, and that the assumption that $S_{Sil}$ is a proxy for dust column in \ion{H}{2} regions is not proven\footnote{It is also (arguably) puzzling that we see dependences on {\itshape any} line deficit at $S_{Sil}\gtrsim1.4$, since (some) models demand that silicate strengths greater than about this value require smooth rather than clumpy dust distributions \citep{nen08}.}. Second, we see significant line deficits for some sources with $S_{Sil}<0$, i.e. a silicate emission feature. Third, only the [\ion{C}{2}] deficit gets stronger with advancing merger stage. If more advanced mergers host dustier \ion{H}{2} regions, then we would also see a dependence of the [\ion{N}{2}] deficit on merger stage. 

These observations suggest that some fraction of the [\ion{C}{2}] deficit is not driven by increased dust in \ion{H}{2} regions. We lack the data to investigate this in detail, so we only briefly discuss this further. We consider three possible origins; (1) increased charging of dust grains \citep[e.g.][]{malho01}, leading to a lower gas heating efficiency, (2) a softer radiation field in the diffuse ISM \citep[e.g.][]{spaans94}, and (3) dense gas in the PDRs, making [\ion{O}{1}] rather than [\ion{C}{2}] the primary coolant. 

The third of these possibilities is feasible, but we do not have the data to confirm or refute it as a mechanism. The second possibility is unlikely, due to the energetic nature of star formation in ULIRGs and from the arguments in \citet{luh03}. For the first possibility; if the origin of the additional deficit in [\ion{C}{2}] is grain charging, then we would see a higher $G_{0}$ in advanced mergers compared to early stage mergers. From Figure \ref{pdrtoolres3} (see \S\ref{pdrtmodel}) the advanced mergers have a roughly order of magnitude higher value of $G_{0}$ for about the same $n$\footnote{We note though that this also holds for the trends discussed with $S_{Sil}$; dividing into two samples with $1.4<S_{Sil}<2.1$ and $S_{Sil}>2.1$ yields a $G_{0}/n$ ratio about a factor of three higher in the latter sample}. We therefore infer, cautiously, and with the caveat that we cannot rule out [\ion{O}{1}] being a major cooling line, that part of the [\ion{C}{2}] deficit arises due to grain charging in PDRs or in the diffuse ISM. We further propose that this increase in grain charge is not driven mainly by a luminous AGN, obscured or otherwise, since we see no dependence of the [\ion{C}{2}] deficit on either optical spectral type or the presence of [\ion{Ne}{5}]14.32.

\begin{deluxetable*}{lccccccc}
\tabletypesize{\scriptsize}
\setlength{\tabcolsep}{0.02in} 
\tablecolumns{8}
\tablewidth{0pc}
\tablecaption{Mid-Infrared Properties and Far-IR derived star formation rates \label{midflux}}
\tablehead{
\colhead{Galaxy}              & \multicolumn{2}{c}{PAH 6.2}     & \multicolumn{2}{c}{PAH 11.2}    &\colhead{S$_{Sil}$}& \colhead{$\langle \rm{SFR}_{\rm{Line}} \rangle$} & \colhead{$\langle \rm{SFR}_{\rm{Cont}} \rangle$} \\
                              & \colhead{Flux}& \colhead{EW}    & \colhead{Flux} &\colhead{EW}    &\colhead{}         & \colhead{M$_{\odot}$yr$^{-1}$}                   & \colhead{M$_{\odot}$yr$^{-1}$}                   
}
\startdata
 \objectname{IRAS 00188-0856} & 1.89$\pm$0.28 & 0.065$\pm$0.009 & 1.45$\pm$0.10  & 0.39$\pm$0.01  &  2.62$\pm$0.01    &  98$\pm$25  &  239$\pm$67   \\
 \objectname{IRAS 00397-1312} & 1.81$\pm$0.28 & 0.026$\pm$0.004 & 0.80$\pm$0.22  & 0.19$\pm$0.06  &  2.94$\pm$0.01    & 655$\pm$207 &  979$\pm$195	\\
 \objectname{IRAS 01003-2238} & 1.02$\pm$0.07 & 0.039$\pm$0.002 & 0.65$\pm$0.14  & 0.02$\pm$0.01  &  0.77$\pm$0.08    & 129$\pm$34  &   97$\pm$14   \\
 \objectname{Mrk 1014}        & 1.91$\pm$0.22 & 0.079$\pm$0.009 & 1.75$\pm$0.12  & 0.07$\pm$0.01  & -0.21$\pm$0.02    & 183$\pm$60  &   -- 		    \\
 \objectname{IRAS 03158+4227} & 1.30$\pm$0.11 & 0.059$\pm$0.005 & 1.19$\pm$0.12  & 0.43$\pm$0.09  &  3.13$\pm$0.01    & 157$\pm$17  &  270$\pm$22	\\
 \objectname{IRAS 03521+0028} & 1.05$\pm$0.07 & 0.355$\pm$0.035 & 0.94$\pm$0.07  & 0.50$\pm$0.01  &  1.29$\pm$0.13    & 137$\pm$28  &  252$\pm$80   \\
 \objectname{IRAS 06035-7102} & 4.42$\pm$0.47 & 0.087$\pm$0.009 & 3.88$\pm$0.27  & 0.28$\pm$0.01  &  1.51$\pm$0.01    & 219$\pm$23  &  138$\pm$24	\\
 \objectname{IRAS 06206-6315} & 2.80$\pm$0.40 & 0.183$\pm$0.026 & 1.96$\pm$0.14  & 0.35$\pm$0.01  &  1.57$\pm$0.16    & 117$\pm$22  &  162$\pm$24	\\
 \objectname{IRAS 07598+6508} & 8.25$\pm$0.27 & 0.006$\pm$0.002 & 0.78$\pm$0.24  & 0.01$\pm$0.005 & -0.13$\pm$0.01    &  75$\pm$16  &   --          \\
 \objectname{IRAS 08311-2459} & 6.38$\pm$0.40 & 0.139$\pm$0.006 & 8.52$\pm$0.30  & 0.06$\pm$0.01  &  0.49$\pm$0.05    & 436$\pm$31  &  190$\pm$21	\\
 \objectname{IRAS 10378+1109} & 9.16$\pm$0.18 & 0.090$\pm$0.017 & 0.70$\pm$0.05  & 0.44$\pm$0.03  &  2.13$\pm$0.01    & 153$\pm$13  &  149$\pm$13	\\
 \objectname{IRAS 11095-0238} & 1.20$\pm$0.50 & 0.037$\pm$0.015 & 1.10$\pm$0.12  & 0.40$\pm$0.07  &  3.28$\pm$0.01    & 131$\pm$21  &  118$\pm$12	\\
 \objectname{IRAS 12071-0444} & 1.66$\pm$0.13 & 0.087$\pm$0.006 & 1.20$\pm$0.01  & 0.08$\pm$0.01  &  1.37$\pm$0.01    & 171$\pm$50  &  143$\pm$16	\\
 \objectname{3C 273}          & $<$1.16       & $<$0.007        & $<$0.32        & $<$0.004       & -0.11$\pm$0.01    & 153$\pm$59  &   56$\pm$19	\\
 \objectname{Mrk 231}         & 6.48$\pm$1.53 & 0.009$\pm$0.002 & 9.02$\pm$0.63  & 0.03$\pm$0.005 &  0.65$\pm$0.07    &  65$\pm$17  &   --	        \\
 \objectname{IRAS 13451+1232} & $<$5.10       & $<$0.02         & $<$0.6         & $<$0.02        &  0.26$\pm$0.03    & 131$\pm$61  &  103$\pm$11	\\
 \objectname{Mrk 463}         & $<$1.04       & $<$0.01         & $<$4.25        & $<$0.04        &  0.41$\pm$0.04    &  90$\pm$27  &   -- 	        \\
 \objectname{IRAS 15462-0450} & 1.79$\pm$0.13 & 0.061$\pm$0.002 & 1.36$\pm$0.11  & 0.07$\pm$0.01  &  0.38$\pm$0.04    & 190$\pm$33  &   --          \\
 \objectname{IRAS 16090-0139} & 2.09$\pm$0.19 & 0.070$\pm$0.006 & 1.74$\pm$0.12  & 0.44$\pm$0.01  &  2.52$\pm$0.01    & 241$\pm$16  &  289$\pm$29	\\
 \objectname{IRAS 19254-7245} & 6.66$\pm$0.93 & 0.066$\pm$0.009 & 4.59$\pm$0.32  & 0.12$\pm$0.01  &  1.33$\pm$0.01    & 165$\pm$20  &   90$\pm$20	\\
 \objectname{IRAS 20087-0308} & 5.37$\pm$0.39 & 0.366$\pm$0.026 & 2.85$\pm$0.20  & 0.70$\pm$0.01  &  1.80$\pm$0.18    & 216$\pm$118 &  319$\pm$13	\\
 \objectname{IRAS 20100-4156} & 2.94$\pm$0.65 & 0.088$\pm$0.019 & 2.12$\pm$0.15  & 0.61$\pm$0.05  &  2.66$\pm$0.01    & 254$\pm$35  &  496$\pm$149	\\
 \objectname{IRAS 20414-1651} & 2.91$\pm$0.20 & 0.570$\pm$0.014 & 1.59$\pm$0.11  & 0.58$\pm$0.01  &  1.61$\pm$0.16    & 106$\pm$12  &   94$\pm$38	\\
 \objectname{IRAS 23230-6926} & 3.27$\pm$0.42 & 0.323$\pm$0.041 & 1.79$\pm$0.12  & 0.54$\pm$0.01  &  2.13$\pm$0.01    & 161$\pm$64  &  139$\pm$16	\\
 \objectname{IRAS 23253-5415} & 1.30$\pm$0.26 & 0.234$\pm$0.068 & 1.25$\pm$0.09  & 0.41$\pm$0.01  &  1.46$\pm$0.15    & 203$\pm$82  &  194$\pm$59	\\   
\enddata 
\tablecomments{Flux units are $\times 10^{-20}$ W cm$^{-2}$. PAH data are taken from the IRS spectra in the CASSIS database (\citealt{lebo11}, see also e.g. \citealt{spo07,desai07}). The PAH 6.2$\mu$m EWs (but not the fluxes) have been corrected for ice absorption. The last two columns give mean star formation rates from lines and continua detected at $>3\sigma$, using Equations \ref{xpahsfr1a}-\ref{xpahsfr1f} and \ref{relations6a}-\ref{relations6f}.}
\end{deluxetable*}

\subsection{Star Formation Rate}\label{sfrdisc}
\subsubsection{Line Luminosities}\label{linessfrdisc}
We examine far-IR fine-structure lines as star formation rate indicators by comparing their luminosities to those of PAHs. PAHs are thought to originate from short-lived Asymptotic Giant Branch stars \citep{gehrz89,habing96,blom05,ber06}, and therefore to be associated with star-forming regions \citep{tiel08}. They are prominent in starburst galaxies but weak in AGNs \citep{lau00,wee05}. While uncertainty remains over how to calibrate PAHs as star formation rate measures \citep{pee04,for04,sargs12}, their luminosities are likely reasonable proxies for the instantaneous rate of star formation. Conversely, for far-IR fine-structure lines the relation between line luminosity and star formation rate is less clear. We may expect a correlation, as the inter-stellar radiation field (ISRF) from young stars may be an important excitation mechanism. Moreover, correlations between far-IR line luminosities and star formation rate have been observed previously \citep[e.g.][]{bos02,delooze11,sargs12,zhao13}. The extent to which correlations exist is however poorly constrained.

With the caveats that PAH luminosity depends both on metallicity \citep{madden06,calz07,khr13} and dust obscuration, neither of which we can correct for, we assume the PAH luminosities give `true' star formation rate measures, which we compare to the far-IR line luminosities. In doing so, we further assume that there is negligible differential extinction between the PAHs and the far-IR lines. To mitigate effects from variation of individual PAH features \citep{jds07}, we sum the luminosities of the PAH 6.2$\mu$m and 11.2$\mu$m features into a single luminosity, L$_{\rm{PAH}}$ \citep{far07}. 

We plot far-IR line luminosities against L$_{\rm{PAH}}$ in Figure \ref{firlinevspahlumfit}. Unlike the plots with L$_{\rm{IR}}$, there is no significant difference between the Sy1s and \ion{H}{2}/LINERs. This is consistent with both PAHs and the far-IR lines primarily tracing star formation. Fitting relations of the form in Equation \ref{baserelation} to all the objects, and converting to star formation rate by using equation 5 of \citet{far07}, yields:

\begin{mathletters}
\begin{eqnarray}
\rm{log}(\dot{M}) & = & ( -7.02\pm1.25) + (1.07\pm0.14)\rm{log}(L_{\rm{[OIII]}})  \label{xpahsfr1a}  \\
                  & = & ( -5.13\pm0.72) + (0.91\pm0.09)\rm{log}(L_{\rm{[NIII]}})  \label{xpahsfr1b}  \\
                  & = & ( -5.44\pm1.79) + (0.86\pm0.20)\rm{log}(L_{\rm{[OI]63}})  \label{xpahsfr1c}  \\
			      & = & ( -7.30\pm0.87) + (1.19\pm0.11)\rm{log}(L_{\rm{[NII]}})   \label{xpahsfr1d}  \\
                  & = & (-10.04\pm1.34) + (1.55\pm0.17)\rm{log}(L_{\rm{[OI]145}}) \label{xpahsfr1e}  \\
                  & = & ( -6.24\pm1.72) + (0.95\pm0.19)\rm{log}(L_{\rm{[CII]}})   \label{xpahsfr1f}    
\end{eqnarray}
\end{mathletters}

\noindent Using the criteria from \S\ref{linelumsdisc} then all the lines show a significant correlation. The trends with [\ion{O}{1}]63 and [\ion{C}{2}] are only barely significant (see also \citealt{diaz13}). Excluding the Sy1s and Sy2s from the fits yields consistent slopes and intercepts in all cases, though the fits are now no longer significant for [\ion{O}{1}]63 and [\ion{C}{2}]. Considering only those lines detected at $>3\sigma$, then the derived star formation rates are, for a given object, consistent to within a factor of three in nearly all cases. We present the mean star formation rates in Table \ref{midflux}. 

An alternative to PAH luminosity as a tracer of star formation rate is the sum of the [\ion{Ne}{3}]15.56$\mu$m and [\ion{Ne}{2}]12.81$\mu$m line luminosities, L$_{\rm{Neon}}$ \citep{thor00,hoketo07,shipley13}. We thus compared L$_{\rm{Neon}}$ to L$_{\rm{Line}}$. For the whole sample, we find relations mostly in agreement with Equations \ref{xpahsfr1a} through \ref{xpahsfr1f}, though the relation with [\ion{C}{2}] is now formally insignificant. Excluding the Seyferts and comparing L$_{\rm{Neon}}$ to L$_{\rm{Line}}$ yields similar results to those with L$_{\rm{PAH}}$.

We note four further points. First, we tested sums of far-IR lines as tracers of star formation rate but found no improvement on the individual relations. Second, if we instead consider L$_{\rm{Line}}/$L$_{\rm{IR}}$ vs. L$_{\rm{PAH}}/$L$_{\rm{IR}}$ then we see correlations with comparable scatter to those in Figure \ref{firlinevspahlumfit}. Third, we investigated whether using L$_{\rm{PAH}}$ is better than using a single PAH luminosity, by reproducing Figure \ref{firlinevspahlumfit} using only the PAH 6.2$\mu$m luminosity. We found consistent relations in all cases, albeit with larger scatter for the L$_{\rm{PAH}}^{6.2}$ plots. Fourth, if we instead code the points in Figure \ref{firlinevspahlumfit} by the 11.2$\mu$m PAH EW then we see no dependence of the relations on the energetic importance of star formation. 

Finally, we note two points about the relation between star formation rate and L$_{\rm{[CII]}}$. First, we investigated whether L$_{\rm{[CII]}}$ shows an improved correlation with star formation rate if objects with a strong [\ion{C}{2}] deficit are excluded. Adopting a (somewhat arbitrary) boundary of L$_{\rm{[CII]}}$/L$_{\rm{IR}}=2\times10^{-4}$ yields only a marginally different relation:

\begin{equation}\label{ciisfrnodef}
\rm{log}(\dot{M}) = ( -6.59\pm1.58) + (0.99\pm0.18)\rm{log}(L_{\rm{[CII]}})   
\end{equation}

\noindent suggesting that the correlation does not depend strongly on the [\ion{C}{2}] deficit, though there is the caveat that star formation rate is derived from L$_{\rm{PAH}}$ (see \S\ref{defdisc} \& Figure \ref{ciilirplots_si}). Second, the relation between $\dot{\rm{M}}_{\odot}$ and L$_{\rm{[CII]}}$ is consistent with that given by \citet{sargs12}, though their sample mostly consists of lower luminosity systems.

\begin{figure*}
\begin{center}
\includegraphics[width=110mm,angle=00]{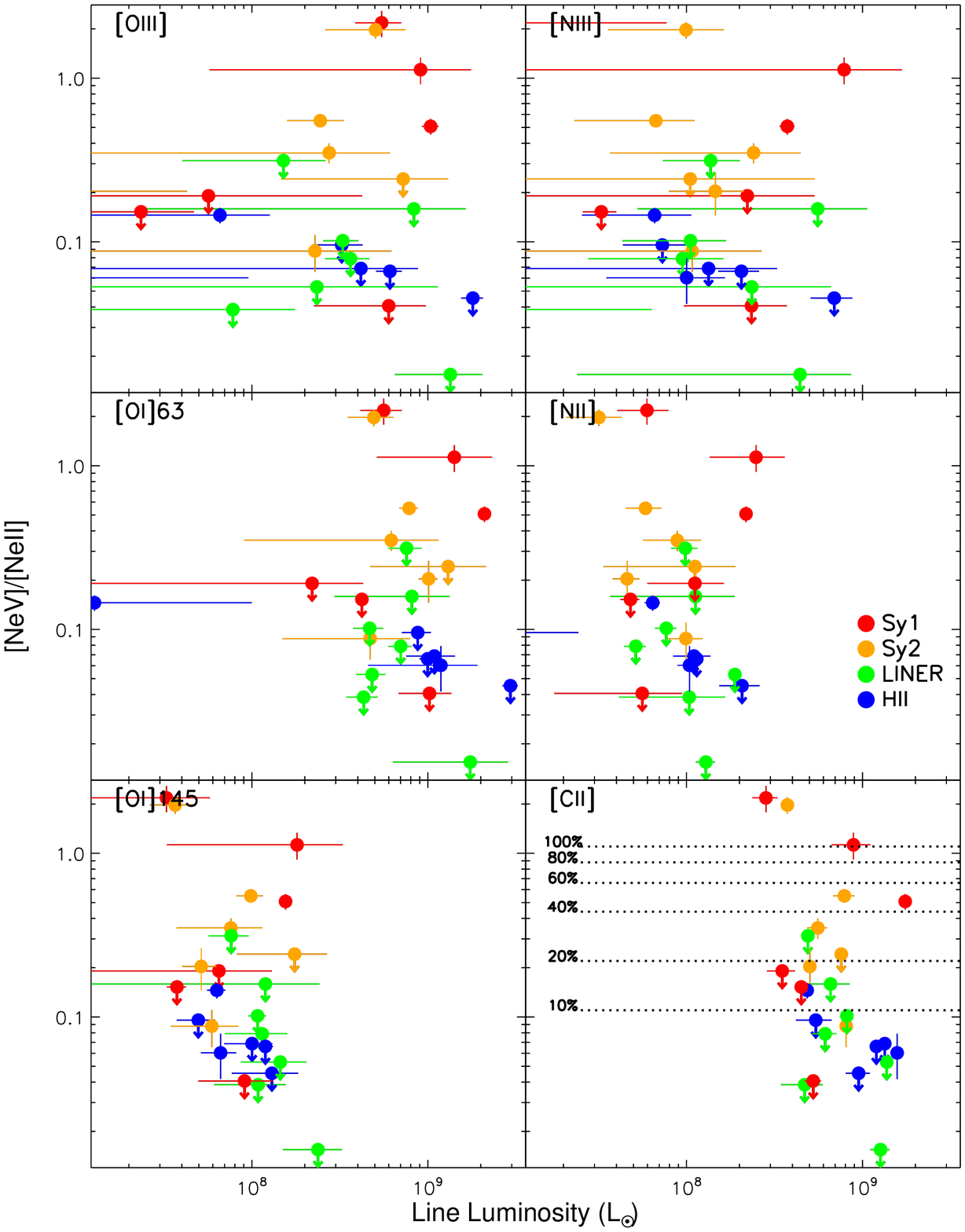}
\caption{Far-IR line luminosities vs [\ion{Ne}{5}]14.32/[\ion{Ne}{2}]12.88 (\S\ref{agnactivity}). The lines in the lower right panel are AGN contribution as a function of [\ion{Ne}{5}]14.32/[\ion{Ne}{2}]12.88, from \citealt{sturm02} (we plot these on only one panel for clarity).}\label{firlinevsnevneii}
\end{center}
\end{figure*}

\subsubsection{Continuum Luminosities}\label{contsfrdisc}
We examine continuum luminosity densities as star formation rate tracers in Figure \ref{contlumssfr}. Employing the same method as in \S\ref{irlumdisc} we find that the continua near all the lines provide acceptable fits. Converting to relations with star formation rate (see \S\ref{sfrdisc}) we find:

\begin{mathletters}
\begin{eqnarray}
\rm{log}(\dot{M}) & = & (3.24\pm0.24) + (1.89\pm0.46)\rm{log}(L_{\rm{52}})  \label{relations6a}  \\
                  & = & (2.95\pm0.10) + (1.38\pm0.19)\rm{log}(L_{\rm{57}})  \label{relations6b}  \\
                  & = & (3.04\pm0.17) + (1.79\pm0.39)\rm{log}(L_{\rm{63}})  \label{relations6c}  \\
				  & = & (2.83\pm0.11) + (0.93\pm0.16)\rm{log}(L_{\rm{122}}) \label{relations6d}  \\
                  & = & (3.02\pm0.18) + (1.09\pm0.25)\rm{log}(L_{\rm{145}}) \label{relations6e}  \\
                  & = & (3.36\pm0.22) + (1.42\pm0.30)\rm{log}(L_{\rm{158}}) \label{relations6f}     
\end{eqnarray}
\end{mathletters}

\noindent which are all significant using the criteria from \S\ref{linelumsdisc}. The correlations with continua at $\leq63\mu$m may however be stronger, consistent with stronger correlations with warmer (star formation heated) dust. This is consistent with findings by previous authors \citep{brandl06,calz07}. We present the mean star formation rates derived from these relations in Table \ref{midflux}. In most cases they are consistent with the line-derived star formation rates.

\subsection{Gas Photoionization}\label{photoio}
If electron densities are below the critical density in the narrow-line region, then the hardness of the radiation field ionizing an element can be estimated via flux ratios of adjacent ionization states of that element; $f_{X^{i+1}}/f_{X^{i}}$. For a fixed $U$ this ratio will be approximately proportional to the number of photons producing the observed $X^{i}$ flux relative to the number of Lyman continuum photons. 

For the mid-IR line-emitting gas, two diagnostic ratios of this type can be used; [\ion{Ne}{3}]15.56/[\ion{Ne}{2}]12.81 and [\ion{S}{4}]10.51/[\ion{S}{3}]18.71. The photon energies required to produce these four ions are all $<50$eV, meaning that they can be produced in star-forming regions \citep{smi01,ber01,pee02,verma03}. As the Neon lines are detected in all of our sample, we use the Neon ratio as a proxy for mid-IR gas excitation\footnote{Though this ratio is unlikely to be a pure star formation tracer due to potential contamination of the [\ion{Ne}{3}]15.56$\mu$m flux by AGN \citep{gorj07}}. We find no trend of this ratio with individual far-IR line luminosities (e.g. Figure \ref{firlinevsneratio}), either for the whole sample or for subsamples divided by optical type or PAH 11.2$\mu$m EW. 

We also examined three mid-to-far and far-IR line ratios to try fashioning an excitation plane diagram, in a similar manner to \citet{dal06}; [\ion{O}{4}]26/[\ion{O}{3}], [\ion{N}{3}]/[\ion{N}{2}], and [\ion{O}{3}]/[\ion{N}{2}]. In no case did we find any trends. The large uncertainties on the [\ion{O}{4}]26, [\ion{O}{3}], and [\ion{N}{3}] lines means though that we cannot conclude that such trends do not exist.

\subsection{AGN Activity}\label{agnactivity}
We first compare far-IR line luminosities to optical spectral classification. We see no trends. If we normalize the line luminosities by L$_{\rm{IR}}$ then no trends emerge, either for individual lines or sums of lines (Figure \ref{firlinevsoptclass}). We also see no trends if we compare optical class to line ratios, or normalized ratios. Moreover, the five objects with an additional broad component in [\ion{C}{2}] (\S\ref{profileres}) do not have an unusually high incidence of Seyfert spectra. We conclude that optical class cannot be inferred from far-IR line luminosities or ratios. This is consistent with the gas producing the optical emission not being strongly associated (in terms of heating mechanism) with the far-IR line emitting gas, at least in the majority of cases. 

\begin{figure}[ht]
\begin{center}
\includegraphics[width=70mm]{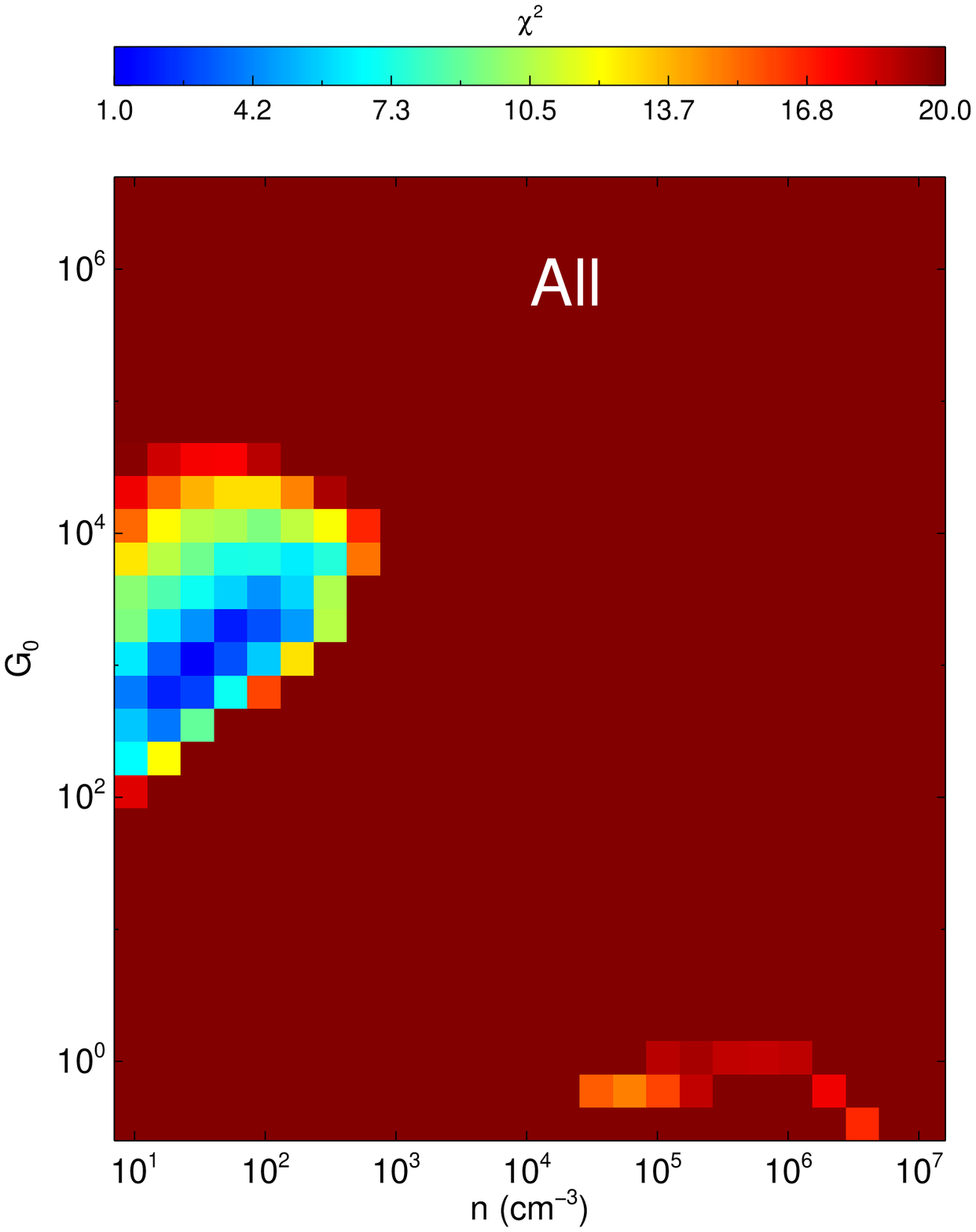}
\caption{Results from simultaneous modelling of the [\ion{O}{1}]63/[\ion{C}{2}], [OI145/CII158], and ([\ion{O}{1}]63 +[\ion{C}{2}])/L$_{\rm{FIR}}$ ratios, using PDRToolbox \citep{kauf06,pou08}, to constrain the electron density $n$ and incident far-UV radiation field intensity $G_{0}$ (\S\ref{pdrtmodel}). The image is the median-combined stacked $n$ vs. $G_{0}$ plane for the whole sample. Units of $n$ are cm$^{-3}$ and units of $G_{0}$ are the local Galactic interstellar FUV field found by \citealt{habing} ($1.6\times10^{-3}$ ergs cm$^{-2}$ s$^{-1})$). The $x$ and $y$ axis ranges are fixed by PDRToolbox.}\label{pdrtoolres}
\end{center}
\end{figure}

Optical spectra may however misclassify AGN in obscured systems as \ion{H}{2} or LINERs. We therefore employ the [\ion{Ne}{5}]14.32/[\ion{Ne}{2}]12.88 line ratio as an AGN diagnostic. Both these lines are less affected by extinction than are optical lines. The [\ion{Ne}{5}]14.32 line can arise in planetary nebulae and supernova remnants \citep{oli99}. For extragalactic sources though, it is weak or absent in star forming regions \citep[e.g.][]{lut98,sturm02,bern09}, but strong in spectra of AGN \citep[e.g.][]{spin09}. The [\ion{Ne}{2}]12.88 line on the other hand is seen almost universally in galaxies. Their ratio should therefore be a reasonable proxy for the presence of an AGN.

We plot [\ion{Ne}{5}]14.32/[\ion{Ne}{2}]12.88 against far-IR line luminosities in Figure \ref{firlinevsnevneii}. There is a correlation between the Neon line ratio and optical classification, but no correlations with far-IR line luminosity. If we substitute optical class for PAH 11.2$\mu$m EW, then no trends emerge. Considering the Sturm et al mixing lines (bottom right panel of Figure \ref{firlinevsnevneii}) then we see no trends among objects classified either as weak or strong AGN. Finally, if we instead plot line luminosity normalized by L$_{\rm{IR}}$ on the $x$ axis, then we still see no trends. 

We searched for trends with far-IR line ratios, normalized ratios, sums and normalized (by L$_{\rm{IR}}$) sums, but found nothing convincing, though the small number of sources with [\ion{Ne}{5}]14.32 detections means we are not certain that no trends exist. We conclude, cautiously, that for ULIRGs there is no reliable diagnostic of AGN luminosity using only simple combinations of far-IR line luminosities. This result is consistent with the weaker correlation observed between L$_{\rm{Line}}$ and L$_{\rm{IR}}$ if Sy1s are included (see \S\ref{linelumsdisc}), if the AGN supplies an effectively random additional contribution to L$_{\rm{IR}}$, thus increasing the scatter in the relation.

\subsection{UV intensity and electron density}\label{pdrtmodel}
From \S\ref{sfrdisc} \& \ref{agnactivity}, it is plausible that at least the plurality of the [\ion{C}{2}] emission arises from PDRs. We defer rigorous modelling to a future paper, and here only estimate the beam-averaged PDR hydrogen nucleus density, $n$ (cm$^{-3}$), and incident far-ultraviolet (FUV; 6 eV $<$ E $<$ 13.6 eV) radiation field intensity $G_{0}$ (in units of the local Galactic interstellar FUV field found by \citealt{habing}; $1.6\times10^{-3}$ erg cm$^{-2}$ s$^{-1}$) using the web-based tool PDR Toolbox\footnote{http://dustem.astro.umd.edu/pdrt/} \citep{kauf06,pou08}. We set constraints using three line ratios; [\ion{O}{1}]63/[\ion{C}{2}], [\ion{O}{1}]145/[\ion{C}{2}], and ([\ion{O}{1}]63 +[\ion{C}{2}])/L$_{\rm{FIR}}$, where L$_{\rm{FIR}}$ is the IR luminosity longward of 30$\mu$m. We assume that all three lines trace PDRs, and that there are no differential extinction effects. We estimate L$_{\rm{FIR}}$ using the same methods as for L$_{\rm{IR}}$ (Table \ref{sample}). 

\begin{figure}[ht]
\begin{minipage}[b]{\linewidth}
\centering
\includegraphics[width=45mm,angle=00]{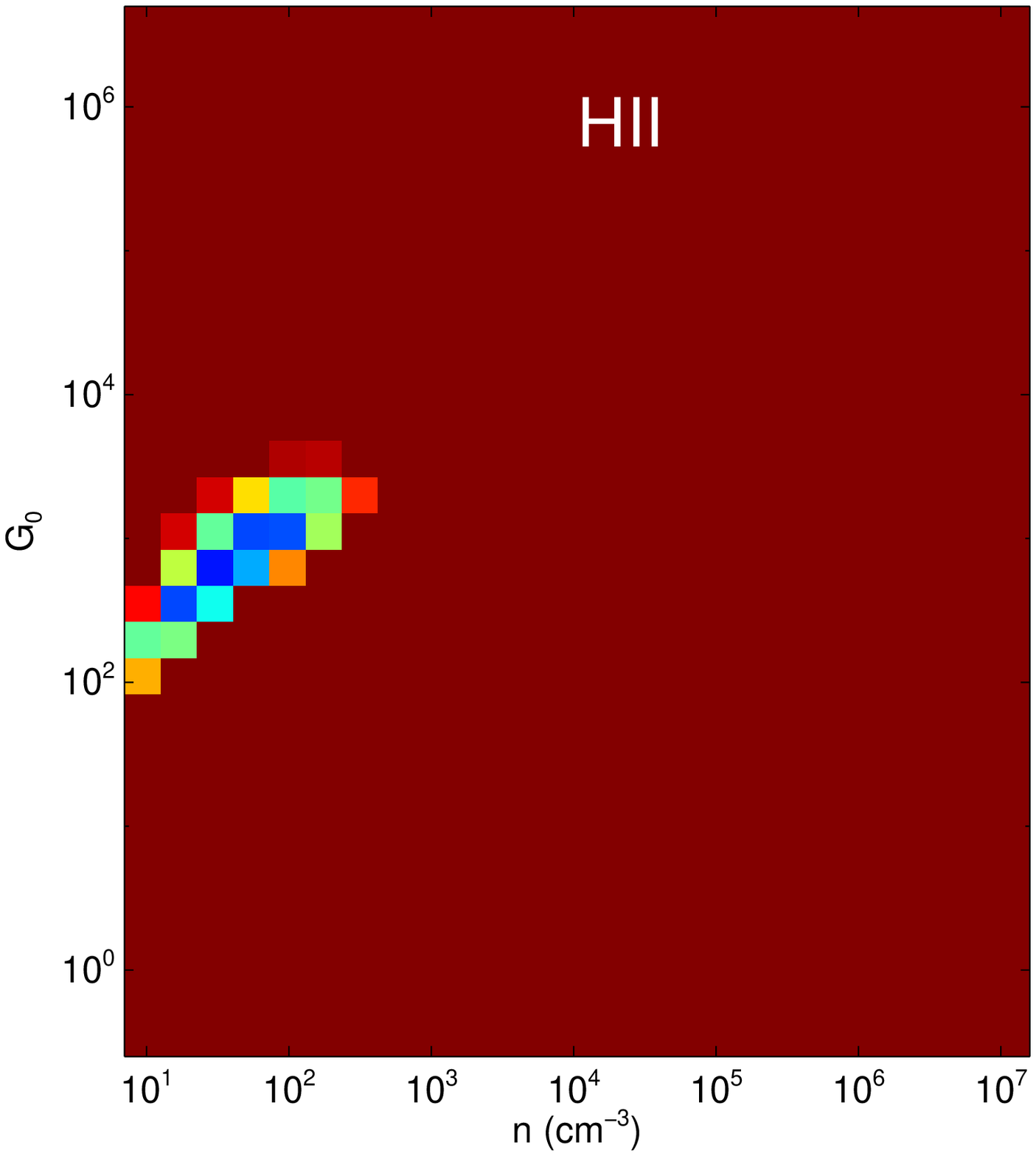} \\
\vspace{-1.2cm}
\includegraphics[width=45mm,angle=00]{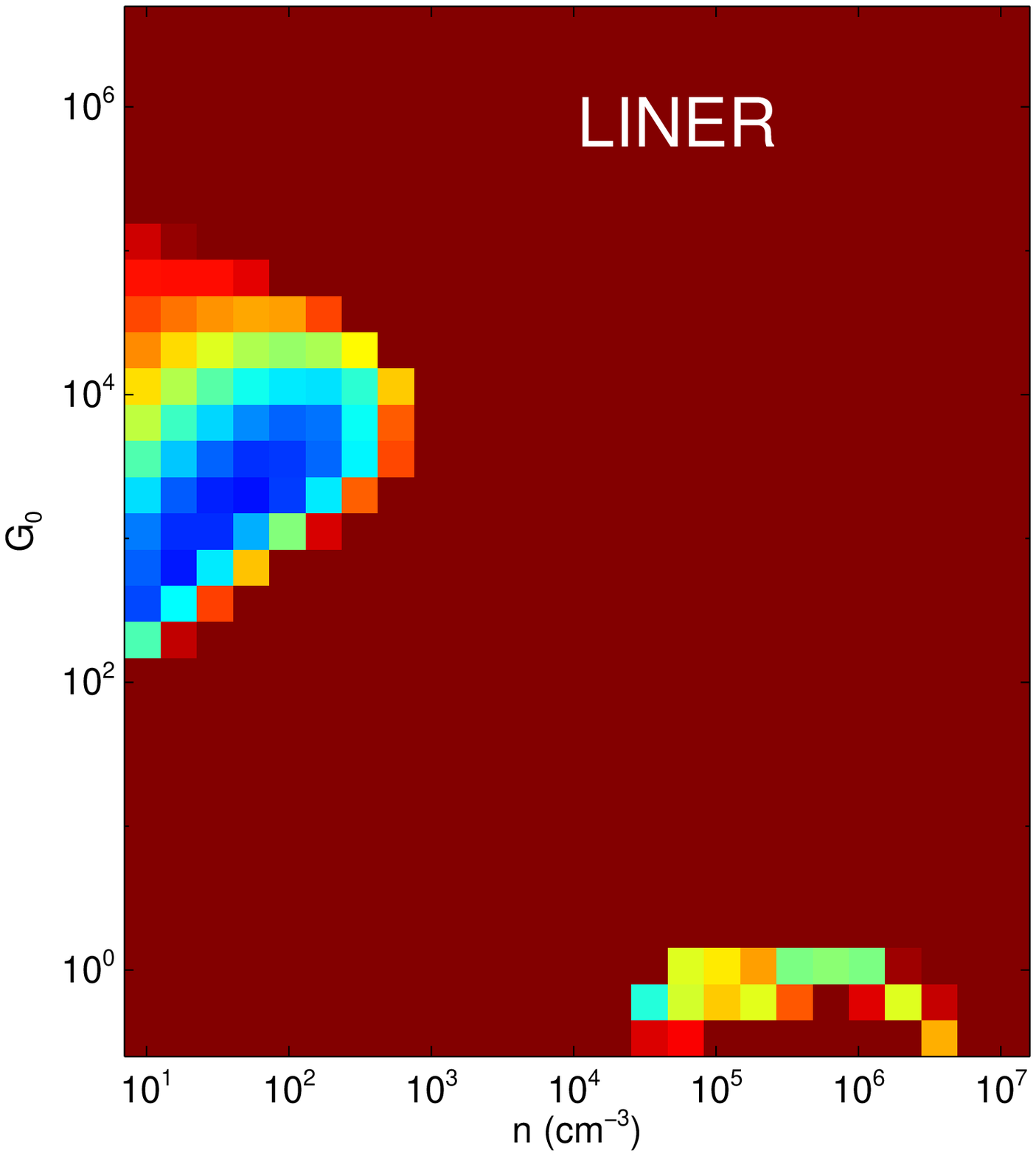} \\
\vspace{-1.2cm}
\includegraphics[width=45mm,angle=00]{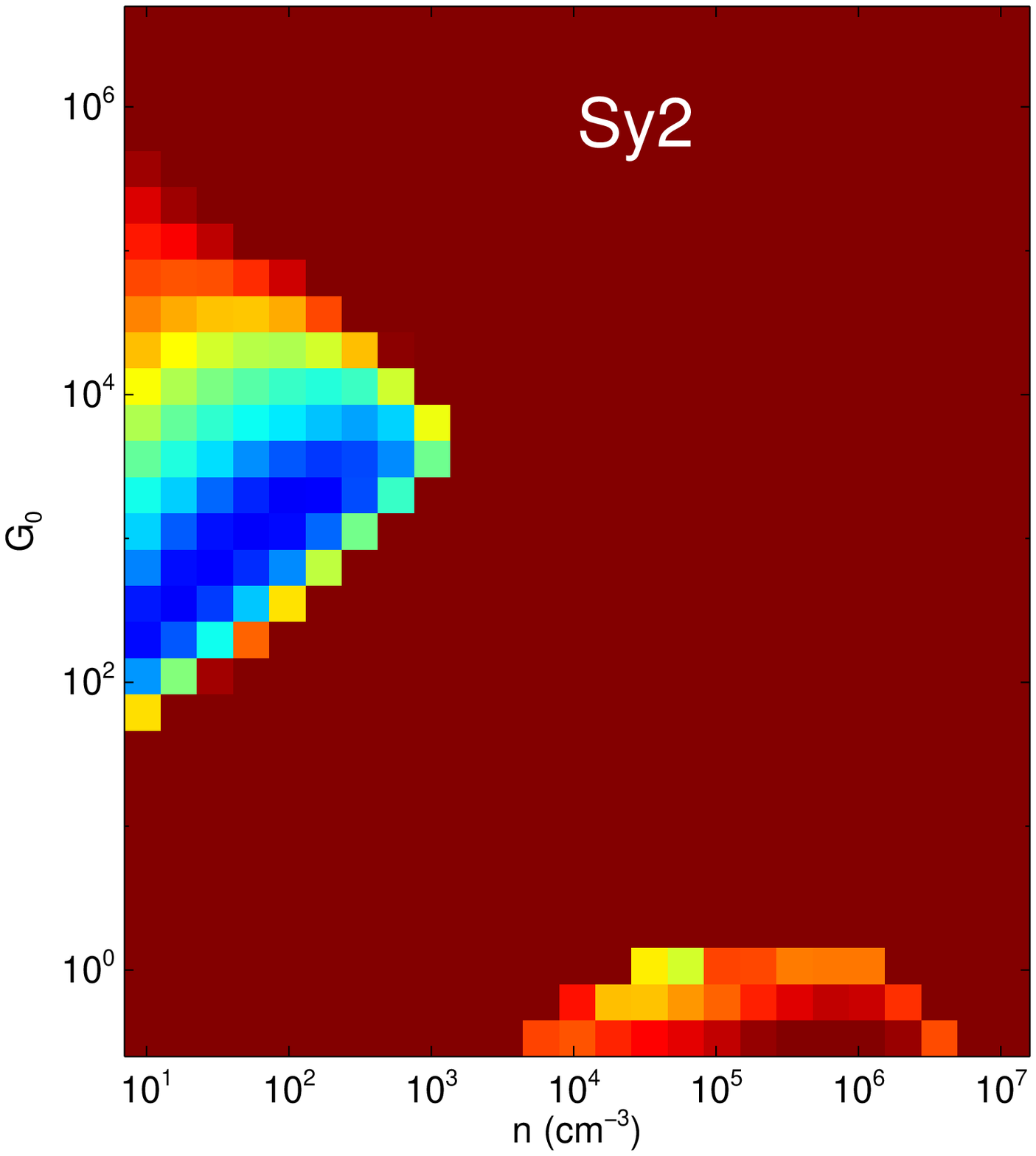} \\
\vspace{-1.2cm}
\includegraphics[width=45mm,angle=00]{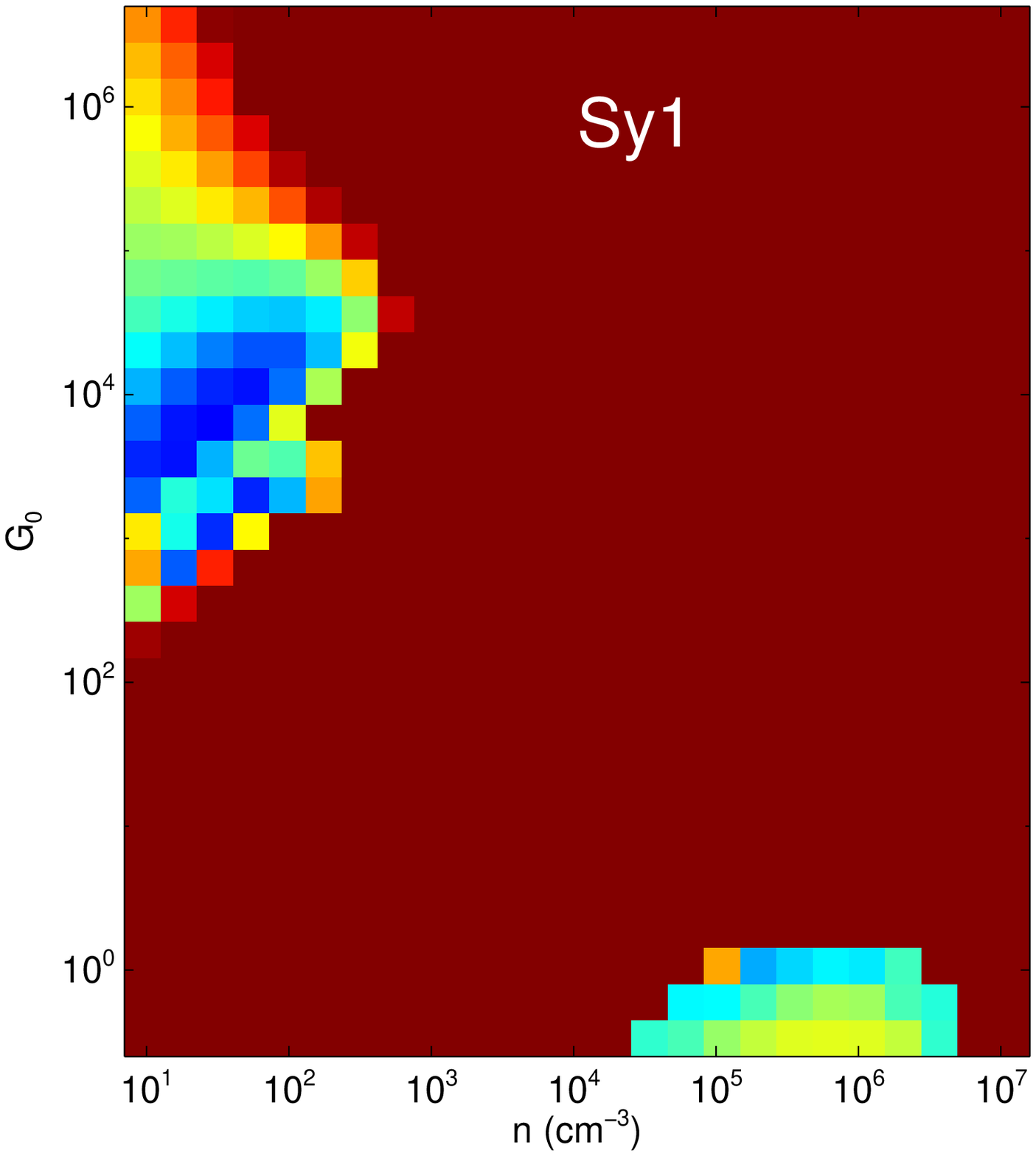} 
\end{minipage}
\caption{Results from PDRToolbox modelling, see Figure \ref{pdrtoolres} for details. The panels show samples divided on optical spectral type.}\label{pdrtoolres2}
\end{figure}

For the whole sample (Figure \ref{pdrtoolres}) we find (taking a conservative cut of $\chi^2_{red}<5$) ranges of $10^{1}<n<10^{2.5}$ and $10^{2.2}<G_{0}<10^{3.6}$, with a power-law dependence between the two. The ranges of both $n$ and $G_{0}$ depend on optical class (Figure \ref{pdrtoolres2}). For \ion{H}{2} objects we find $10^{1.1}<n<10^{2.2}$ and $10^{2.4}<G_{0}<10^{3.3}$. For LINERS and Seyferts however the ranges widen; for LINERS we find $10^{0.8}<n<10^{2.5}$ and $10^{2.4}<G_{0}<10^{4.1}$, and for Sy2s we find $10^{0.7}<n<10^{3}$ and $10^{1.9}<G_{0}<10^{3.9}$. For Sy1s the range for $n$ is comparable to that of LINERs and Sy2s but the range for $G_{0}$ increases to $10^{2.5}<G_{0}<10^{4.7}$. For Sy1s there is a secondary solution that is close to acceptable, which has $G_{0}$ and $n$ values approximately four orders of magnitude lower and higher, respectively, than the primary solution. 

\begin{figure}[ht]
\begin{minipage}[b]{\linewidth}
\centering
\includegraphics[width=45mm,angle=00]{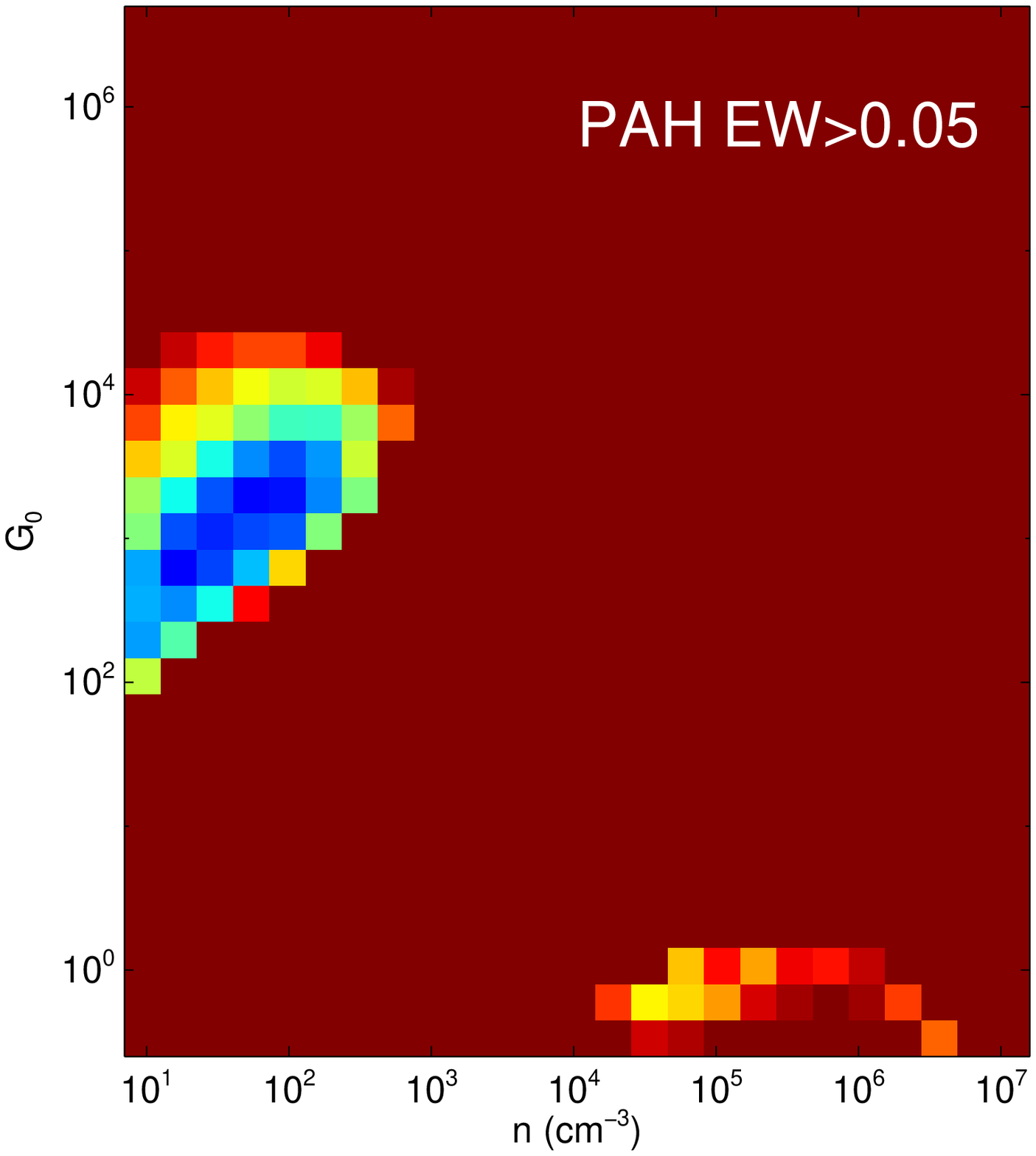} \\
\vspace{-1.2cm}
\includegraphics[width=45mm,angle=00]{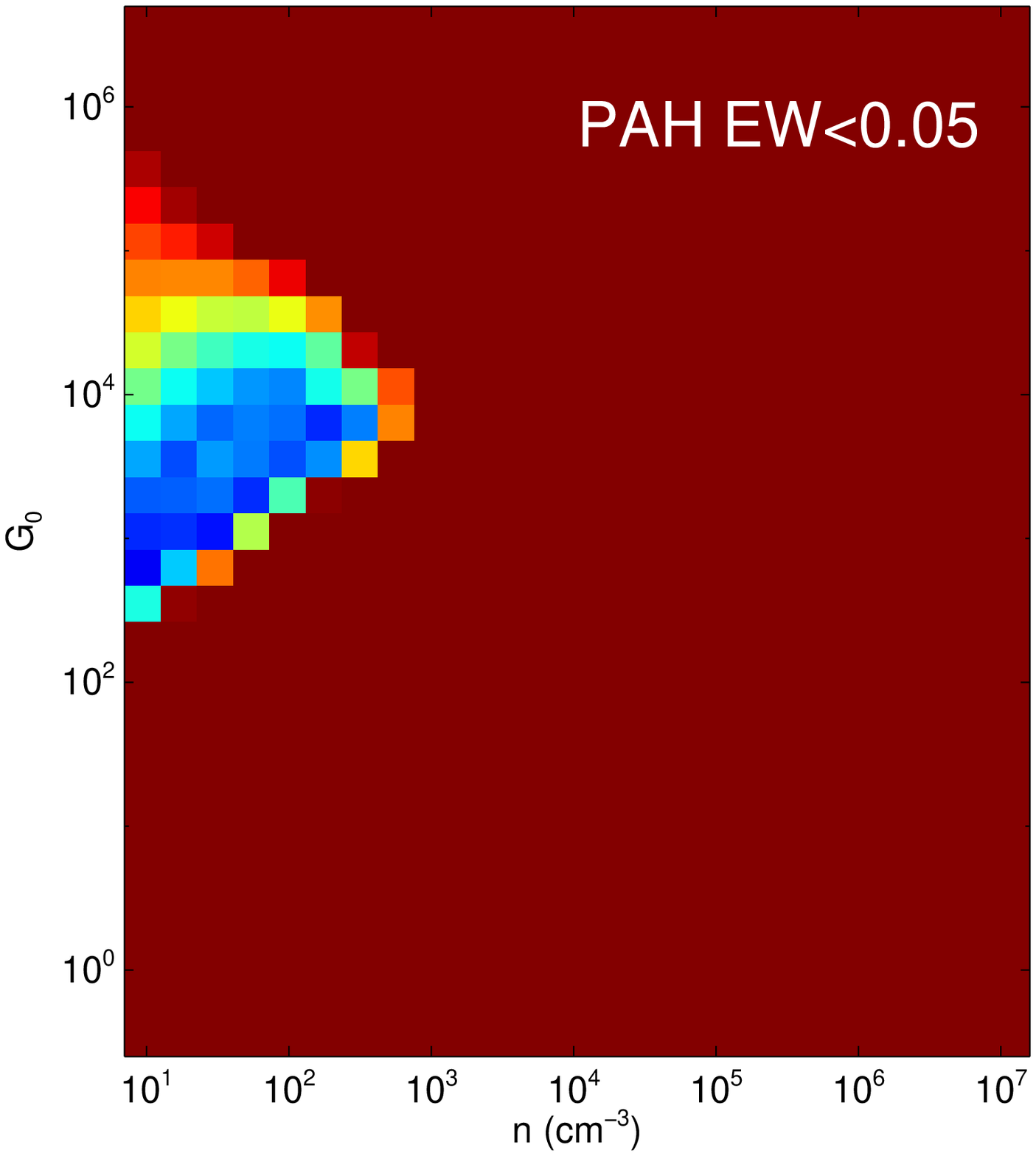} \\
\vspace{-1.2cm}
\includegraphics[width=45mm,angle=00]{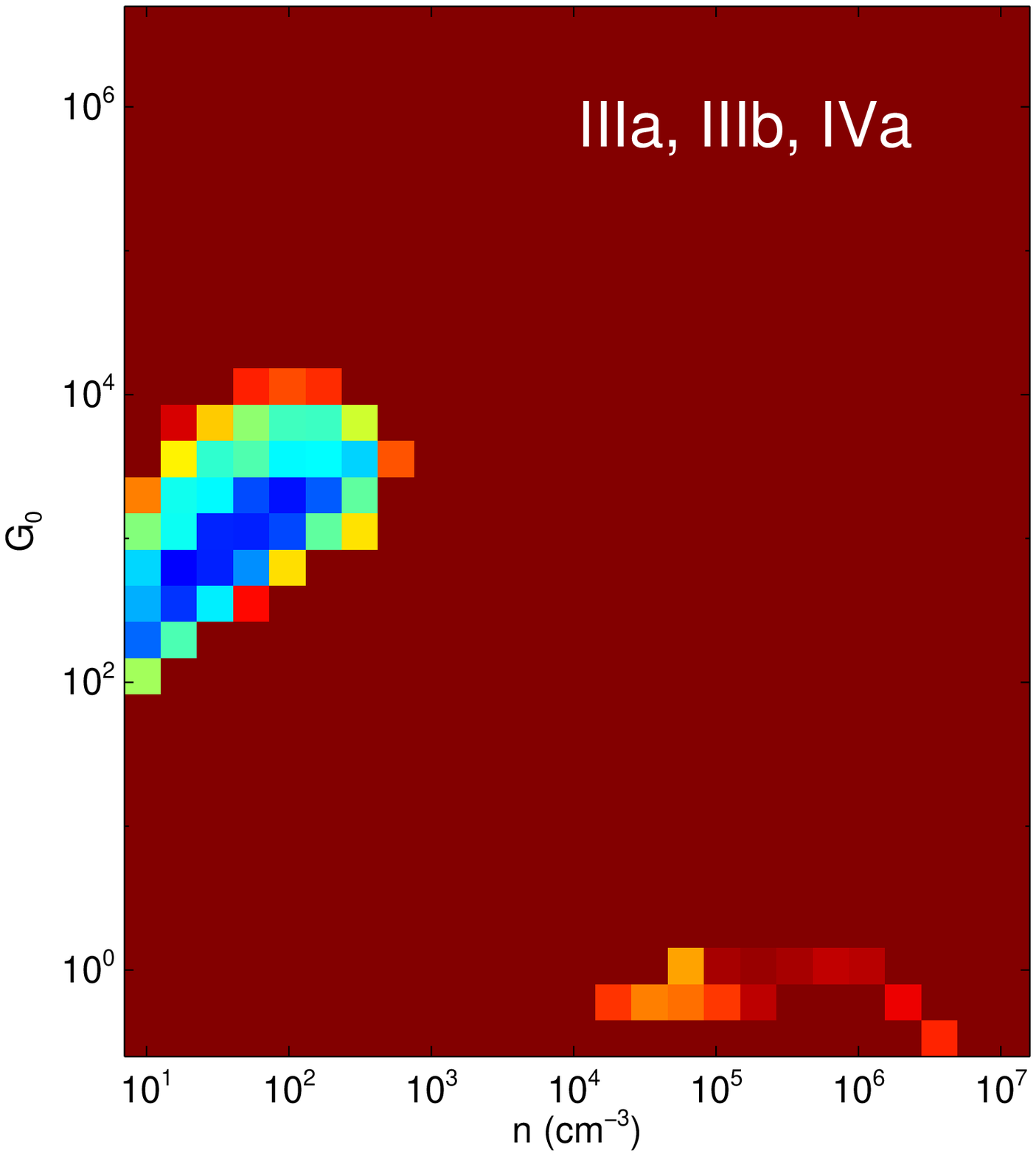} \\
\vspace{-1.2cm}
\includegraphics[width=45mm,angle=00]{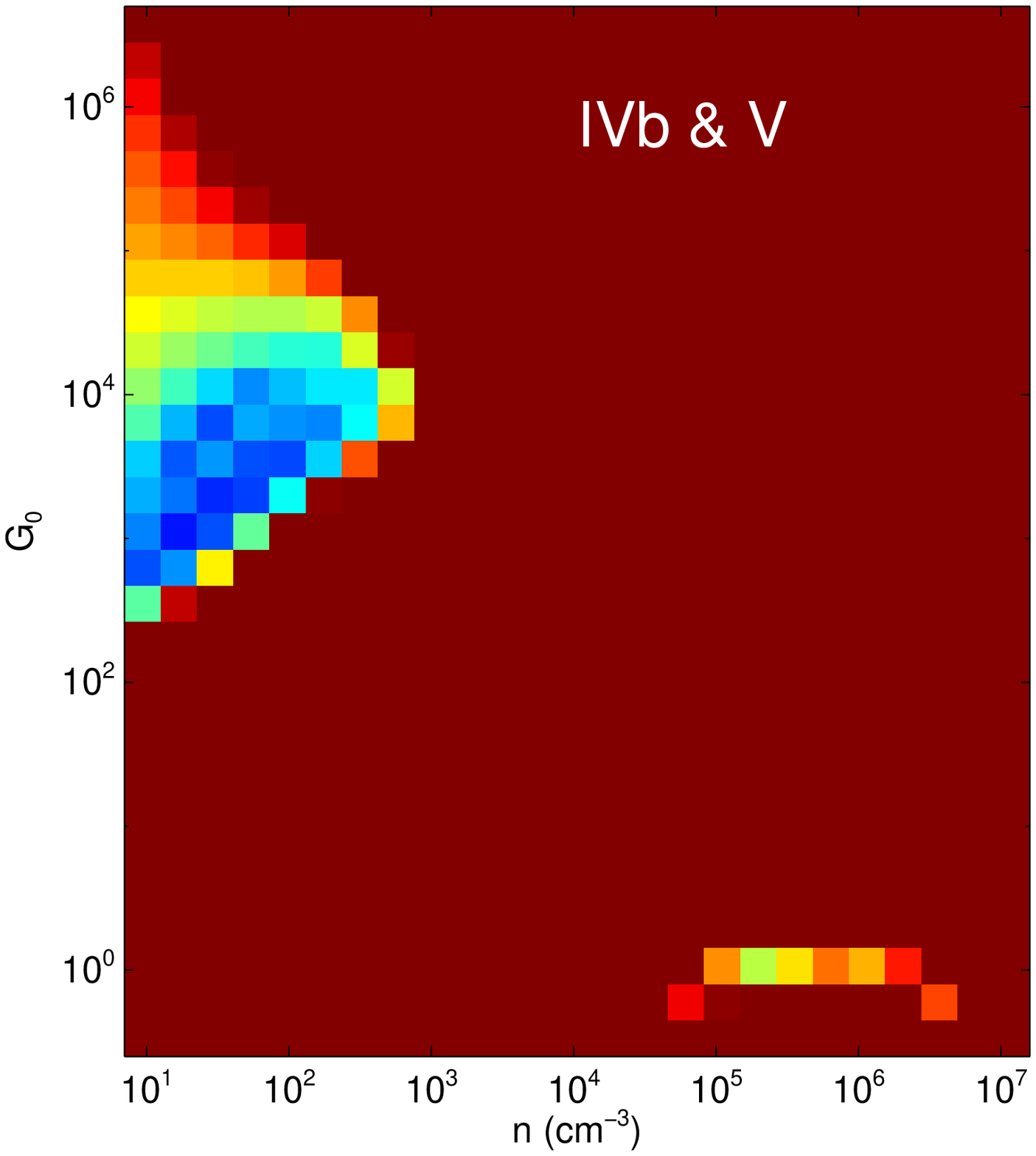} 
\end{minipage}
\caption{Results from PDRToolbox modelling, see Figure \ref{pdrtoolres} for details. The panels show samples divided on PAH 11.2$\mu$m EW ($<0.05\mu$m vs. $\geq0.05\mu$m) and merger stage (Table \ref{sample}).}\label{pdrtoolres3}
\end{figure}

If we divide the sample in two by PAH 11.2 EW (top row of Figure \ref{pdrtoolres3}) then we see a difference, also. The range in $n$ for both samples is comparable, at about $10^{0.8}<n<10^{2.5}$. The ranges for $G_{0}$ are however different; for objects with prominent PAHs it is $10^{2.1}<G_{0}<10^{3.7}$ while for objects with weak PAHs it is $10^{2.6}<G_{0}<10^{4.3}$. We obtain similar ranges for both parameters if we instead divide the sample on merger stage (bottom row of Figure \ref{pdrtoolres3}). This is consistent with a more intense ISRF destroying PAH molecules (see also e.g. \citealt{hern09}). It is however also consistent with a luminous AGN (with a harder UV radiation field) arising after the star formation has faded. This would give the same observation but with no direct relation between the two phenomena.

There are three caveats in using these models to estimate $G_{0}$ and $n$ for our sample. First, we cannot account for different beam filling factors for different lines. This is potentially a significant problem for [\ion{C}{2}] (see \S\ref{lumincorrs} and \S\ref{defdisc}). Second, these models have difficulty in predicting PAH emission strengths \citep{luh03,abel09}, suggesting an incomplete description of the dependence of far-IR line strengths on dust-grain size distribution, PAH properties and ISRF spectral shape (see also e.g. \citealt{okada13}). Since our targets are dusty, it is surprising that we obtain reasonable solutions, indicating that the line ratios and the adopted IR luminosities are compatible with each other. Our derived parameter ranges for $G_{0}$ and $n$ should however be viewed with caution. 

Third is that the [\ion{O}{1}] lines are complex to model. Assuming emission in PDRs then, like [\ion{C}{2}], the [\ion{O}{1}] lines are expected to form within $A_v \sim 3$ magnitudes of the PDR surfaces. It is in these regions that all of the carbon and oxygen should be ionized and atomic, respectively, with gas temperatures between about 250 and 700K \citep[e.g.][]{kauf99}. The dust in these regions has only a small effect on [\ion{C}{2}], but can have a large impact on the [\ion{O}{1}] levels, which are affected by both radiative and collisional processes. Such processes can alter the power of [\ion{O}{1}]63 via the absorption of 63$\mu$m line-emitted photons by dust grains, or by pumping of oxygen atoms by 63$\mu$m continuum dust emitted photons. The effect of dust should not be neglected when modeling [\ion{O}{1}]63, or the [\ion{O}{1}] line ratio, in sources that are optically thick at wavelengths shorter than 100$\mu$m \citep{gonz08}.

\subsection{Merger Stage}\label{mrgedisc}
\begin{figure*}
\begin{center}
\includegraphics[width=110mm,angle=00]{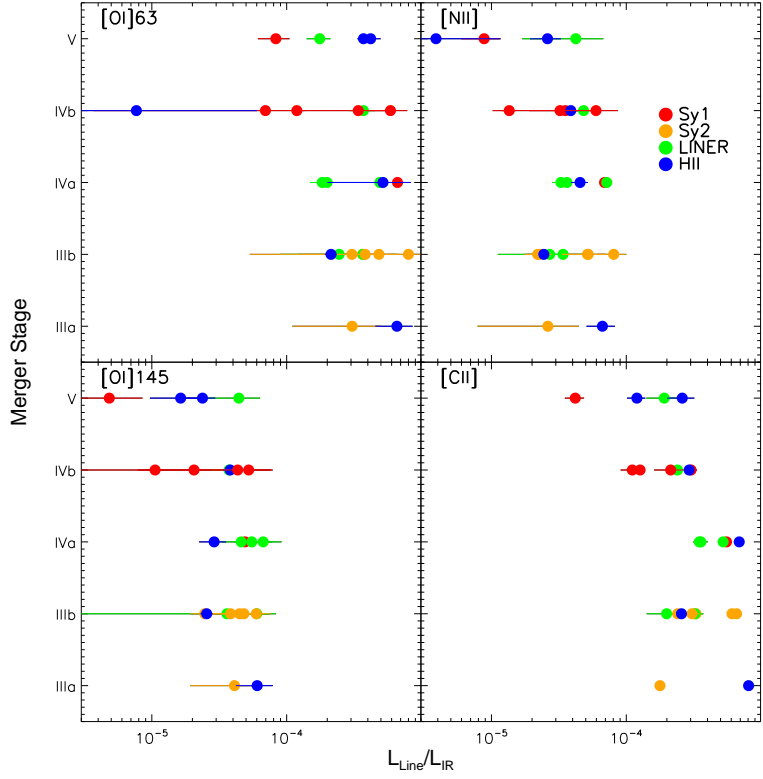}
\caption{Line luminosities normalized by L$_{\rm{IR}}$, vs Merger Stage (\S\ref{mrgedisc}).}\label{merge1}
\end{center}
\end{figure*}

There is evidence that the power source in ULIRGs evolves as a function of merger stage, with star formation dominating, on average, until the progenitors coalesce, whereupon an AGN sometimes becomes energetically important \citep[e.g.][]{rig99,far09}. We may therefore see a correlation between far-IR line properties and merger stage. We find however no correlations between merger stage and far-IR line luminosities. We also see no trend with any far-IR line ratio. Comparing merger stage to normalized far-IR line luminosities (Figure \ref{merge1}) there may be a weak trend; for [\ion{O}{1}]63 and longer lines, the advanced mergers {\itshape might} show a smaller normalized line luminosity than earlier stage mergers. This is consistent with the line luminosities tracing star formation, and with star formation becoming less important as merger stage advances. However, the trend is not strong, and does not depend on optical classification or PAH 11.2$\mu$m EW.

\subsection{SMBH Mass}\label{smbhdisc}
Scaling relations have been derived between the masses of supermassive black holes and the FWHM and continuum luminosities of several UV, optical, and mid-IR emission lines \citep{kasp00,vest02,vest06,das08}. We here explore, using SMBH masses derived from optical lines (Table \ref{sample}), whether there exist correlations between far-IR line properties and SMBH mass. While the absolute uncertainties on the SMBH masses from these studies are of order 0.4 dex, the relative uncertainties within the sample are likely smaller as we focus on one class of object and use H$\beta$ derived masses in nearly all cases. We therefore assume an error on SMBH mass of 20$\%$. 

\begin{figure*}
\begin{center}
\includegraphics[width=110mm,angle=00]{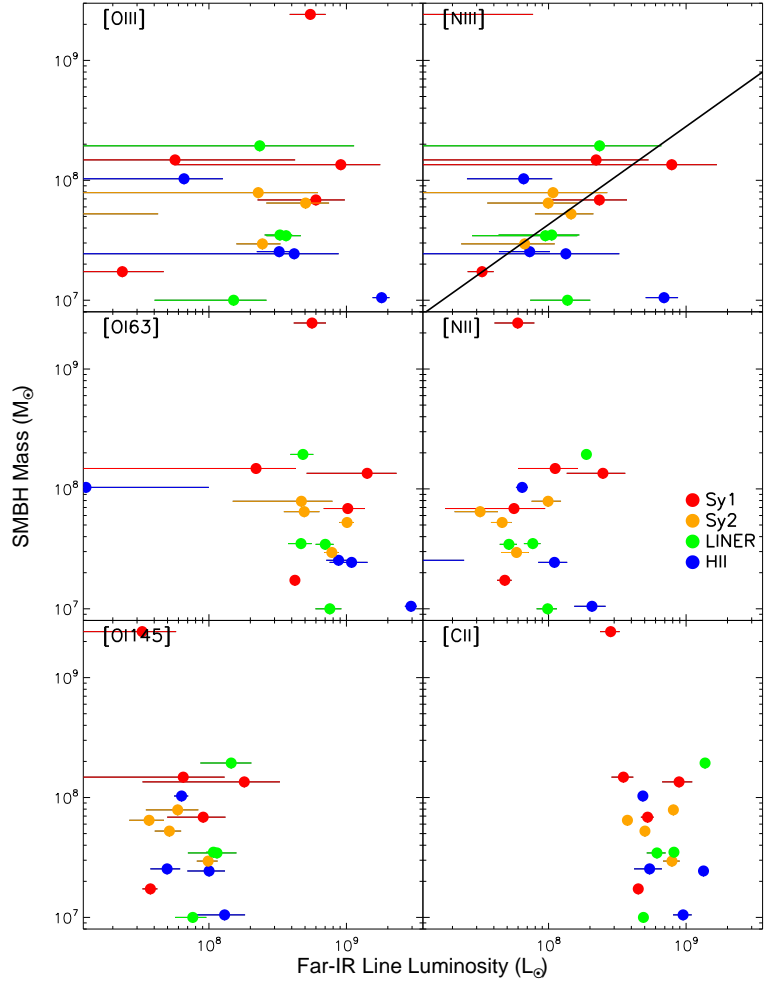}
\caption{Far-IR line luminosities vs SMBH Mass (\S\ref{smbhdisc}). The fit in the [\ion{N}{3}] panel is Equation \ref{smbhmass}, and is to the Seyferts, excluding 3C273.}\label{smbhmassinvestigate}
\end{center}
\end{figure*}

We compare the SMBH masses to line luminosities in Figure \ref{smbhmassinvestigate}. For the whole sample, no line shows a trend with SMBH mass. Considering only the Sy1 and Sy2s, and excluding 3C273, then some line luminosities show, qualitatively, a positive trend with SMBH mass. The trend is only significant for L$_{\rm{[NIII]}}$, for which we derive:

\begin{equation}\label{smbhmass}
\rm{log}(M_{\rm{BH}}) =  (1.09\pm1.43)  +  (0.82\pm0.18) \rm{log}(L_{\rm{[NIII]}})
\end{equation}

\noindent We see similar results if we instead compare M$_{\rm{SMBH}}/$L$_{\rm{IR}}$ to L$_{\rm{Line}}/$L$_{\rm{IR}}$.

It is plausible to exclude 3C273, since it is the only Blazar in the sample. We do not, however, claim that this relation is real, for four reasons. First, if we assume a (still reasonable) error on the SMBH masses in excess of 30\% then the relation is no longer significant. Second, there is no trend of L$_{\rm{[NIII]}}$ with the AGN diagnostics considered in \S\ref{agnactivity}. Third, if this relation is real then it is strange that we do not see a correlation of SMBH mass with L$_{\rm{[OIII]}}$ (see Table \ref{linesobs}, though there is a potentially important difference; [\ion{N}{3}] is a ground-state transition whereas [\ion{O}{3}] is not). Fourth, we searched for correlations between SMBH mass and far-IR continuum luminosities near $57\mu$m, but did not find any clear relations.

\section{Conclusions}
We have presented observations with PACS onboard Herschel of 25 ULIRGs at $z<0.27$. We observed each ULIRG in six lines: [\ion{O}{3}]52$\mu$m,  [\ion{N}{3}]57$\mu$m, [\ion{O}{1}]63$\mu$m, [\ion{N}{2}]122$\mu$m, [\ion{O}{1}]145$\mu$m, and [\ion{C}{2}]158$\mu$m. We used the properties of these lines, together with diagnostics at other wavelengths, to draw the following conclusions:

1 - In most cases the line profiles are reproducible by single gaussians, with widths between 250 km s$^{-1}$ and 600 km s$^{-1}$. The exceptions are [\ion{O}{1}]63 and [\ion{C}{2}], which occasionally show self absorption and a second, broad component, respectively. We do not see significant systemic offsets of the far-IR lines compared to the optical redshifts. The line luminosities range from just under $10^7$L$_{\odot}$ to just over $2\times10^9$L$_{\odot}$. The [\ion{O}{1}]63 and [\ion{C}{2}] lines are usually the most luminous, while [\ion{O}{1}]145 and [\ion{N}{2}] are usually the least luminous. The line luminosities correlate with each other, though in no case is the correlation particularly strong. Simple line ratio diagnostics suggest relatively low gas densities, on average, and that a significant fraction of the [\ion{C}{2}] emission originates from outside \ion{H}{2} regions. 

2 -  There is a deficit in the [\ion{O}{1}]63/L$_{\rm{IR}}$, [\ion{N}{2}]/L$_{\rm{IR}}$, [\ion{O}{1}]145/L$_{\rm{IR}}$, and [\ion{C}{2}]/L$_{\rm{IR}}$ ratios compared to lower luminosity systems, of factors of 2.75, 4.46, 1.50, and 4.95, respectively. There is evidence that the [\ion{C}{2}] and [\ion{N}{2}] deficits correlate with 9.7$\mu$m silicate feature strength (S$_{Sil}$); if S$_{Sil}\gtrsim1.4$ then the [\ion{C}{2}] and [\ion{N}{2}] deficits rise with increasing S$_{Sil}$. We also see a correlation between [L$_{\rm{PAH}}$]/L$_{\rm{IR}}$ and S$_{Sil}$. Furthermore, the [\ion{C}{2}] deficit correlates with merger stage; objects in advanced mergers show a greater deficit than objects in early stage mergers. These results are consistent with the majority of the line deficits arising due to increased levels of dust in \ion{H}{2} regions. We propose though that a significant fraction of the [\ion{C}{2}] deficit arises from an additional mechanism, plausibly grain charging in PDRs and/or the diffuse ISM. 

3 - The line luminosities only weakly correlate with IR luminosity. The correlations improve if Sy1 objects are excluded. Doing so, and fitting a relation of the form $\rm{log}(L_{\rm{IR}}) = \alpha + \beta\rm{log}(L_{\rm{Line}})$, yields:

\begin{eqnarray}
\rm{log}(L_{\rm{IR}}) & = & (4.46\pm1.77) + (0.92\pm0.20)\rm{log}(L_{\rm{[OIII]}})  \\
                      & = & (4.74\pm1.59) + (0.94\pm0.19)\rm{log}(L_{\rm{[NIII]}})  \\
                      & = & (4.28\pm1.89) + (0.91\pm0.21)\rm{log}(L_{\rm{[OI]63}})  \\
					  & = & (5.29\pm1.57) + (0.89\pm0.20)\rm{log}(L_{\rm{[NII]}})   \\
                      & = & (1.75\pm2.11) + (1.34\pm0.27)\rm{log}(L_{\rm{[OI]145}}) \\
                      & = & (6.73\pm2.44) + (0.64\pm0.27)\rm{log}(L_{\rm{[CII]}})      
\end{eqnarray}

\noindent The best tracers of L$_{\rm{IR}}$ are thus the five shorter wavelength lines. The [\ion{C}{2}] line is a poor tracer of L$_{\rm{IR}}$, accurate to about an order of magnitude at best. Its accuracy does not noticeably improve if objects with a strong [\ion{C}{2}] deficit are excluded. 

4 - The continuum luminosity densities near the wavelengths of the lines correlate with L$_{\rm{IR}}$, irrespective of the presence of Sy1s. We derive:

\begin{eqnarray}
\rm{log}(L_{\rm{IR}}) & = & (12.87\pm0.09) + (0.88\pm0.17)\rm{log}(L_{\rm{52}})  \\
                      & = & (12.86\pm0.05) + (0.91\pm0.11)\rm{log}(L_{\rm{57}})   \\
                      & = & (12.86\pm0.08) + (0.98\pm0.18)\rm{log}(L_{\rm{63}})   \\
					  & = & (12.95\pm0.07) + (0.87\pm0.11)\rm{log}(L_{\rm{122}})  \\
                      & = & (13.05\pm0.11) + (0.90\pm0.16)\rm{log}(L_{\rm{145}})  \\
                      & = & (13.08\pm0.11) + (0.83\pm0.11)\rm{log}(L_{\rm{158}})    
\end{eqnarray}

5 - We find correlations between star formation rate, estimated using L$_{\rm{PAH}}$, and both line luminosities and continuum luminosity densities. For line luminosities we derive:

\begin{eqnarray}
\rm{log}(\dot{M}_{\odot}) & = & ( -7.02\pm1.25) + (1.07\pm0.14)\rm{log}(L_{\rm{[OIII]}})   \\
                          & = & ( -5.13\pm0.72) + (0.91\pm0.09)\rm{log}(L_{\rm{[NIII]}})   \\
                          & = & ( -5.44\pm1.79) + (0.86\pm0.20)\rm{log}(L_{\rm{[OI]63}})   \\
						  & = & ( -7.30\pm0.87) + (1.19\pm0.11)\rm{log}(L_{\rm{[NII]}})    \\
                          & = & (-10.04\pm1.34) + (1.55\pm0.17)\rm{log}(L_{\rm{[OI]145}})  \\
                          & = & ( -6.24\pm1.72) + (0.95\pm0.19)\rm{log}(L_{\rm{[CII]}})      
\end{eqnarray}

\noindent while for the continuum luminosity densities we derive:

\begin{eqnarray}
\rm{log}(\dot{M}_{\odot}) & = & (3.24\pm0.24) + (1.89\pm0.46)\rm{log}(L_{\rm{52}})   \\
                          & = & (2.95\pm0.10) + (1.38\pm0.19)\rm{log}(L_{\rm{57}})   \\
                          & = & (3.04\pm0.17) + (1.79\pm0.39)\rm{log}(L_{\rm{63}})   \\
						  & = & (2.83\pm0.11) + (0.93\pm0.16)\rm{log}(L_{\rm{122}})  \\
                          & = & (3.02\pm0.18) + (1.09\pm0.25)\rm{log}(L_{\rm{145}})  \\
                          & = & (3.36\pm0.22) + (1.42\pm0.30)\rm{log}(L_{\rm{158}})     
\end{eqnarray}

\noindent On average, the shorter wavelength continua show stronger correlations. 

6 - Assuming the [\ion{O}{1}] and [\ion{C}{2}] lines arise mainly in PDRs, we use a simple model to extract estimates for the hydrogen nucleus density, $n$, and incident far-ultraviolet radiation field $G_{0}$, in the far-IR line emitting gas. We find $10^{1}<n<10^{2.5}$ and $10^{2.2}<G_{0}<10^{3.6}$ for the whole sample, with a power-law dependence between the two. The ranges depend on optical spectral class; for \ion{H}{2}-like objects we find $10^{1.1}<n<10^{2.2}$ and $10^{2.4}<G_{0}<10^{3.3}$, while for Sy1s we find $10^{0.8}<n<10^{2.7}$ and $10^{2.5}<G_{0}<10^{4.7}$. There is also a dependence of $G_{0}$ on the {\itshape importance} of star formation; objects with weak PAHs have $10^{2.6}<G_{0}<10^{4.3}$ while objects with prominent PAHs have $10^{2.1}<G_{0}<10^{3.7}$. We find similar ranges for early- vs. late-stage mergers. This is consistent with, but not exclusively supportive of, a more intense ISRF destroying PAH molecules. 

7 - We searched for relations between far-IR line luminosities and ratios, and several other parameters; AGN activity (either from optical spectral class or the detection of [\ion{Ne}{5}]14.32), merger stage, mid-IR excitation, and SMBH mass. For the first three parameters we found no relations. We conclude that the far-IR lines do not arise primarily due to AGN activity, and that the properties of the far-IR line emitting gas do not strongly depend on either mid-IR excitation or merger stage. For SMBH mass we found one superficially striking correlation, with $L_{\rm{[NIII]}}$, but subsequent tests were not supportive. We conclude that far-IR line luminosities do not straightforwardly trace SMBH mass.

\acknowledgments
We thank the staff of the Herschel helpdesk for many valuable discussions, and the referee for a very helpful report. Herschel is an ESA space observatory with science instruments provided by European-led Principal Investigator consortia and with important participation from NASA. This work is based on observations made with the Spitzer Space Telescope. Spitzer is operated by the Jet Propulsion Laboratory, California Institute of Technology under a contract with NASA. This research has made extensive use of the NASA/IPAC Extragalactic Database (NED) which is operated by the Jet Propulsion Laboratory, California Institute of Technology, under contract with NASA, and of NASA's Astrophysics Data System. This research has also made use of Ned Wrights online cosmology calculator \citep{wright06}. V.L. is supported by a CEA/Marie Curie Eurotalents fellowship. J.A. acknowledges support from the Science and Technology Foundation (FCT, Portugal) through the research grants PTDC/CTE-AST/105287/2008, PEst-OE/FIS/UI2751/2011 and PTDC/FIS-AST/2194/2012. E.G-A is a Research Associate at the Harvard-Smithsonian Center for Astrophysics, and thanks the support by the Spanish Ministerio de Econom\'{\i}a y Competitividad under projects AYA2010-21697-C05-0 and FIS2012-39162-C06-01.

{\it Facilities:} \facility{Herschel}, \facility{Spitzer}.

\begin{turnpage}

\begin{deluxetable*}{lcccccccccccccccccc}
\tabletypesize{\scriptsize}
\tablecolumns{19}
\tablewidth{0pc}
\tablecaption{Far-Infrared line Fluxes \label{linefluxc}}
\tablehead{
\colhead{Galaxy}         &\multicolumn{3}{c}{[O III]}        &\multicolumn{3}{c}{[N III]}      &\multicolumn{3}{c}{[O I]63}        &\multicolumn{3}{c}{[N II]}        &\multicolumn{3}{c}{[O I]145}      &\multicolumn{3}{c}{[C II]}  
}
\startdata
00188-0856  	         &  0.64$\pm$0.80 & 350$\pm$349 &(1) &       $<$0.52 &350$\pm$159 &(1) &  3.54$\pm$0.71\tablenotemark{c} & 225$\pm$36  &(1) & 0.85$\pm$0.51 &296$\pm$38  & (2) &  0.89$\pm$0.39 &350$\pm$53  &(2) &  3.85$\pm$1.02                  &279$\pm$13 &(2) \\
00397-1312  	         &  3.05$\pm$0.43 & 259$\pm$40  &(1) & 1.17$\pm$0.31 &372$\pm$90  &(1) &  4.99$\pm$0.49                  & 328$\pm$30  &(1) & 0.35$\pm$0.09 &333$\pm$85  & (1) &  0.22$\pm$0.09 &250$\pm$94  &(1) &  1.61$\pm$0.25\tablenotemark{a} &427$\pm$61 &    \\
01003-2238  	         &  3.20$\pm$1.00 & 400$\pm$99  &(1) & 0.72$\pm$0.29 &150$\pm$48  &(1) &  8.65$\pm$1.61                  & 197$\pm$34  &(1) & 0.08$\pm$0.16\tablenotemark{e} &400$\pm$639 & (1) &  0.49$\pm$0.12 &165$\pm$39  &(1) &  5.36$\pm$1.23\tablenotemark{d} &150$\pm$7  &(2) \\
Mrk 1014         	     &  4.43$\pm$4.15 & 270$\pm$59  &(2) & 3.83$\pm$4.33 &250$\pm$253 &(2) &  6.89$\pm$4.39 				  & 350$\pm$41  &(2) & 1.21$\pm$0.55 &507$\pm$50  & (2) &  0.88$\pm$0.72 &358$\pm$69  &(2) &  4.33$\pm$1.07                  &316$\pm$17 &(2) \\
03158+4227  	         &  5.38$\pm$4.29 & 300$\pm$178 &(2) & 0.78$\pm$3.20 &150$\pm$87  &(2) &  9.69$\pm$6.20 				  & 155$\pm$47  &(2) & 0.83$\pm$0.58 &150$\pm$45  & (2) &  1.30$\pm$0.69 &214$\pm$48  &(2) &  5.63$\pm$0.11                  &243$\pm$4  &(2) \\
03521+0028  	         &  4.74$\pm$4.60 & 350$\pm$232 &(2) & 3.17$\pm$2.87 &350$\pm$169 &(2) &  4.62$\pm$2.94 				  & 500$\pm$145 &(2) & 0.64$\pm$0.43 &351$\pm$102 & (2) &  0.68$\pm$0.70 &350$\pm$114 &(2) &  3.75$\pm$1.08                  &442$\pm$27 &(2) \\
06035-7102     	         &  9.52$\pm$10.53& 215$\pm$105 &(2) & 3.05$\pm$4.42 &400$\pm$134 &(2) & 24.89$\pm$7.66 				  & 322$\pm$16  &(2) & 2.52$\pm$0.60 &373$\pm$21  & (2) &  2.29$\pm$0.70 &242$\pm$16  &(2) & 30.58$\pm$2.04\tablenotemark{d} &277$\pm$2  &(4) \\
06206-6315  	         &  4.56$\pm$5.54 & 350$\pm$247 &(2) & 3.99$\pm$3.38 &350$\pm$146 &(2) & 10.27$\pm$8.77\tablenotemark{c} & 350$\pm$170 &(2) & 1.47$\pm$0.53 &418$\pm$45  & (2) &  1.26$\pm$0.64 &450$\pm$66  &(2) &  9.25$\pm$1.19                  &369$\pm$10 &(2) \\
07598+6508  	         &  0.34$\pm$2.20 & 150$\pm$127 &(2) & 1.33$\pm$1.87 &150$\pm$237 &(2) &  1.32$\pm$1.25 				  & 153$\pm$101 &(0) & 0.67$\pm$0.31 &282$\pm$60  & (2) &  0.39$\pm$0.39 &184$\pm$74  &(1) &  2.10$\pm$0.38                  &197$\pm$35 &(1) \\
08311-2459  	         & 14.42$\pm$1.54 & 379$\pm$32  &(1) & 5.20$\pm$0.50 &263$\pm$20  &(1) & 29.16$\pm$1.08 				  & 306$\pm$9   &(1) & 3.03$\pm$0.15 &334$\pm$14  & (1) &  2.16$\pm$0.15 &348$\pm$20  &(1) & 24.32$\pm$0.35                  &272$\pm$3  &(1) \\
10378+1109  	         &  1.09$\pm$0.80 & 351$\pm$211 &(1) & 0.99$\pm$0.46 &350$\pm$131 &(1) &  5.46$\pm$1.17 				  & 466$\pm$83  &(1) & 0.71$\pm$0.12 &439$\pm$64  & (1) &  0.55$\pm$0.14 &475$\pm$101 &(1) &  3.53$\pm$0.20                  &409$\pm$20 &(1) \\
11095-0238  	         &  4.45$\pm$1.25 & 323$\pm$73  &(1) & 1.16$\pm$0.82 &356$\pm$202 &(1) &  8.58$\pm$1.28\tablenotemark{c} & 343$\pm$41  &(1) & 0.63$\pm$0.09 &375$\pm$46  & (1) &  1.40$\pm$0.54 &358$\pm$43  &(2) &  7.51$\pm$1.18\tablenotemark{d} &296$\pm$9  &(2) \\
12071-0444  	         &  2.01$\pm$0.71 & 300$\pm$85  &(1) & 0.55$\pm$0.36 &225$\pm$118 &(1) &  6.43$\pm$0.78 				  & 225$\pm$22  &(1) & 0.48$\pm$0.11 &225$\pm$42  & (1) &  0.81$\pm$0.14 &295$\pm$43  &(1) &  6.48$\pm$0.90\tablenotemark{d} &250$\pm$6  &(2) \\
3C 273                   &  2.84$\pm$0.83 & 461$\pm$111 &(1) &       $<$0.40 &300$\pm$400 &(1) &  2.92$\pm$0.77 				  & 418$\pm$92  &(1) & 0.31$\pm$0.10 &500$\pm$134 & (1) &  0.17$\pm$0.13 &500$\pm$321 &(1) &  1.47$\pm$0.24                  &465$\pm$66 &(1) \\
Mrk 231\tablenotemark{b} &  --            & --          &    & 2.80$\pm$0.60 &177         &    & 36.00$\pm$2.60 				  & 218         &    & 4.10$\pm$0.50 &266         &     &  3.20$\pm$0.40 &208         &    & 38.30$\pm$1.30                  &247        &    \\
13451+1232		         &        $<$0.38 & 300$\pm$101 &(1) & 1.30$\pm$0.59 &350$\pm$154 &(1) &  9.01$\pm$1.10 				  & 467$\pm$46  &(1) & 0.41$\pm$0.07 &300$\pm$42  & (1) &  0.46$\pm$0.10 &300$\pm$53  &(1) &  4.48$\pm$0.21                  &465$\pm$18 &(1) \\
Mrk 463        	         & 28.73$\pm$13.84& 272$\pm$23  &(2) & 5.66$\pm$3.62 &369$\pm$70  &(2) & 28.09$\pm$8.13 				  & 338$\pm$15  &(2) & 1.81$\pm$0.63 &425$\pm$27  & (2) &  2.09$\pm$0.59 &250$\pm$11  &(2) & 21.33$\pm$0.88\tablenotemark{d} &250$\pm$2  &(2) \\
15462-0450		     	 &  8.45$\pm$5.27 & 125$\pm$17  &(2) & 3.30$\pm$1.94 &350$\pm$76  &(2) & 14.41$\pm$4.78 				  & 258$\pm$19  &(2) & 0.79$\pm$0.54 &125$\pm$14  & (2) &  1.28$\pm$0.58 &252$\pm$35  &(2) &  7.40$\pm$0.79                  &163$\pm$4  &(4) \\
16090-0139		         & 10.11$\pm$5.21 & 500$\pm$104 &(2) & 3.33$\pm$3.15 &370$\pm$156 &(2) & 13.13$\pm$8.36 				  & 357$\pm$40  &(2) & 0.97$\pm$0.12 &464$\pm$50  & (2) &  1.79$\pm$0.66 &437$\pm$48  &(2) &  9.54$\pm$1.20                  &352$\pm$8  &(2) \\
19254-7245		         &  8.86$\pm$15.21& 475$\pm$211 &(2) & 4.18$\pm$6.19 &475$\pm$642 &(2) & 18.24$\pm$12.46\tablenotemark{c}& 475$\pm$89  &(2) & 3.85$\pm$0.93 &600$\pm$38  & (2) &  2.30$\pm$0.95 &520$\pm$59  &(2) & 31.37$\pm$1.92                  &514$\pm$12 &(2) \\
20087-0308		         &  2.92$\pm$11.26& 475$\pm$315 &(2) & 2.93$\pm$5.32 &550$\pm$362 &(2) &  6.02$\pm$1.14 				  & 511$\pm$72  &(1) & 2.35$\pm$0.10 &550$\pm$18  & (1) &  1.81$\pm$0.73 &489$\pm$55  &(2) & 17.14$\pm$0.18                  &550$\pm$5  &(2) \\
20100-4156		         &  4.90$\pm$0.81 & 202$\pm$27  &(1) & 1.65$\pm$0.43 &300$\pm$64  &(1) &  8.02$\pm$0.81 				  & 235$\pm$19  &(1) & 0.92$\pm$0.08 &300$\pm$23  & (1) &  0.96$\pm$0.10 &300$\pm$27  &(1) &  9.65$\pm$0.19\tablenotemark{d} &247$\pm$4  &(1) \\
20414-1651		         &  1.24$\pm$1.14 & 550$\pm$389 &(1) & 1.24$\pm$0.76 &475$\pm$223 &(1) &  0.24$\pm$1.64\tablenotemark{c} & 550$\pm$895 &(1) & 1.21$\pm$0.12 &550$\pm$42  & (1) &  1.19$\pm$0.14 &550$\pm$50  &(1) &  9.13$\pm$0.46                  &475$\pm$12 &(2) \\
23230-6926		         &  4.02$\pm$0.91 & 250$\pm$44  &(1) & 1.29$\pm$0.76 &375$\pm$173 &(1) &  5.72$\pm$1.12 				  & 288$\pm$46  &(1) & 0.94$\pm$0.13 &375$\pm$43  & (1) &  1.32$\pm$0.15 &279$\pm$27  &(1) &  9.99$\pm$0.20                  &307$\pm$5  &(1) \\
23253-5415		         &        $<$0.76 & 375$\pm$648 &(1) & 0.80$\pm$0.52 &375$\pm$194 &(1) &  9.47$\pm$5.81 				  & 503$\pm$88  &(2) & 0.83$\pm$0.12 &411$\pm$49  & (1) &  0.53$\pm$0.12 &375$\pm$69  &(1) & 12.57$\pm$1.08\tablenotemark{d} &473$\pm$8  &(2) \\
\enddata 
\tablenotetext{a}{SPIRE-FTS flux, see \S\ref{measmeth}.}
\tablenotetext{b}{Measurements taken from \citealt{fisch10}.}
\tablenotetext{c}{May be self-absorbed, see \S\ref{profileres}.}
\tablenotetext{d}{May show asymmetry or an additional broad component, see \S\ref{profileres}.}
\tablenotetext{e}{May be unusually low, see \S\ref{lumincorrs}.}
\tablecomments{For each line, first the flux is given in units of $\times 10^{-21}$ W cm$^{-2}$, then the line width in km s$^{-1}$, then in brackets the method used to measure the flux (\S\ref{measmeth}).}
\end{deluxetable*}
\end{turnpage}

\end{document}